\documentclass[prb,twocolumn,amsmath,amssymb]{revtex4}

\usepackage{color}
\usepackage{bm}
\usepackage{dcolumn}
\usepackage[dvips]{graphicx}
\usepackage{slashed}

\usepackage{ulem}

\newcommand{\ket}[1]{|#1\rangle}

\newcommand{\braket}[2]{\langle#1|#2\rangle}
\newcommand{\tr}{{\rm tr}\,}

\begin{document}

\preprint{preprint}

\title{
Topological magnetoelectric pump in three dimensions
}

\author{Takahiro Fukui}
\author{Takanori Fujiwara}
\affiliation{Department of Physics, Ibaraki University, Mito 310-8512, Japan}

\date{\today}

\begin{abstract}
We study the topological pump for a lattice fermion model mainly in three spatial dimensions.
We first calculate the U(1)  current density
for the Dirac model defined in continuous space-time to review the known results
as well as to introduce some technical details convenient for the calculations of the lattice model.
We next investigate the U(1) current density for a lattice fermion model, a variant of the Wilson-Dirac model.
The model we introduce is defined on a lattice in space but in continuous time,
which is suited for the study of the topological pump.
For such a model, we derive the conserved U(1) current density and  calculate 
it directly for the $1+1$ dimensional system as well as $3+1$ dimensional system 
in the limit of the small lattice constant.
We find that the current includes a nontrivial lattice effect characterized by the Chern number, 
and therefore, the pumped particle number is quantized by the topological reason.
Finally we study the topological temporal pump in $3+1$ dimensions by numerical calculations.
We discuss the relationship between the second Chern number and the first Chern number, the bulk-edge correspondence, 
and the generalized Streda formula which enables us to compute the second Chern number using the spectral asymmetry.
\end{abstract}

\pacs{
}

\maketitle

\section{Introduction}

In a topological background such as a soliton or a vortex, 
the vacuum state of the Dirac fermion shows a nontrivial topological structure.
\cite{Goldstone:1981aa,Wilczek:1987ab}
One of famous examples in condensed matter physics 
is  the mid-gap states of the SSH soliton,\cite{Su:1979aa,Su:1980aa}
which can be effectively described by a Dirac fermion model with a nontrivial mass term.\cite{Takayama:1980aa}
Recent discovery of topological insulators \cite{Kane:2005aa,Qi:2008aa}
tells us that topological states of matter are richer than we expected, \cite{Hasan:2010fk,Qi:2011kx}
and the Dirac fermion model is very convenient for the classification of symmetry classes.
\cite{Schnyder:2008aa,Ryu:2010uq}

The topological pump in one dimensional (1D) systems has been proposed by Thouless,\cite{Thouless:1983fk}
and  experimentally observed quite recently. \cite{Nakajima:2016aa,Lohse:2016aa}
This can also be described very simply by the Dirac fermion, as already seen in Ref. \cite{Goldstone:1981aa}.
The topological pump  has been generalized to three dimensional (3D) systems.\cite{Qi:2008aa}
The 3D pump is unique, since it is a part of topological magneto-electric effect, \cite{Qi:2008aa}
which has close relationship with the chiral anomaly of the Dirac fermion.\cite{Wilczek:1987ab}
Here, the pumping parameter plays a role of the axion field.\cite{Wilczek:1987ab}
The magneto-electric response in generic systems has also been studied by developing
the theory of the orbital magneto-electric polarization. 
\cite{Essin:2009aa,Essin:2010aa,Malashevich:2010aa}

The chiral magnetic effect (CME), originally proposed for the quark-gluon plasma, \cite{Fukushima:2008aa}
has also been attracting much interest in condensed matter physics. 
In particular, the discovery of the Weyl semimetal 
\cite{Murakami:2007aa,Burkov:2011ab,Wan:2011aa,Xiong:2015aa,1503.08179,Zhang:2017aa}
has led our interest to the observation of the chiral anomaly in a crystal through
the magneto-electric response,  \cite{Nielsen:1983aa,Chen:2013aa,Parameswaran:2014aa} including
the CME, \cite{Basar:2014aa,Zyuzin:2012aa,Jian-Hui:2013aa,Chang:2015aa,Ma:2015aa,Sumiyoshi:2016ab,Alavirad:2016aa,OBrien:2017aa,
Sekine:2016aa,Sekine:2016ab}
the anomalous Hall effect (AHE),
\cite{Yang:2011aa,Zyuzin:2012aa,Grushin:2012aa,Vazifeh:2013aa,Goswami:2013aa,Sekine:2016aa,Sekine:2016ab}
axial-magneto-electric effect, \cite{Grushin:2016aa,Pikulin:2016aa} and
Z$_2$ anomaly in Dirac semimetals,  \cite{Burkov:2016aa} etc. 
Thus, the chiral anomaly and its related phenomena have been one of hot topics in condensed matter physics.

In this paper, we examine mainly the topological pump in a 3D system using a lattice fermion model.
In the next section \ref{s:D}, we present the U(1) current of the Dirac fermion in continuous space-time. 
Here, some notations are fixed and some techniques convenient to the lattice model are given.
In Sec. \ref{s:WD_la}, we introduce a Wilson-Dirac model defined on the spatial 1D or 3D lattice 
but in continuous time to study the topological temporal pump in Sec. \ref{s:MEP}. 
Throughout the paper, this model is simply referred to as Wilson-Dirac model.
In Sec. \ref{s:WD_cc}, we derive the conserved U(1) current density for the Wilson-Dirac model.
Because of continuous time, the charge density is the same as that of the Dirac fermion, whereas the current density
includes some lattice effects.
Therefore, 
we calculate the charge density and the current density separately in Sec. \ref{s:ChaDen} and Appendix \ref{s:CurDen}.
In Sec. \ref{s:E}, we calculate the Chern numbers exactly. In particular, we show that the 3D model has nontrivial 
second Chern number $1$ or $-2$. It should be noted that such second Chern numbers are due to the Berry curvature
of the wave function in the zero field limit. 
Derivation of the exact conserved current and the second Chern number for the lattice Wilson-Dirac model 
are one of main results of the present paper.

In Sec. \ref{s:MEP},  we restrict our discussions to the temporal 3D pump, and present some numerical results in detail.
Various numerical analyses of the 3D pump which give physical interpretations of the exact results in Sec. \ref{s:WD}
are another main results of the present paper.
In Sec. \ref{s:3dpump},  the number of pumped particles is explicitly derived in the case of a static and uniform 
magnetic field. This number is proportional to the second Chern number as well  as the magnetic field and 
the system size.\cite{Qi:2008aa}
In 3D systems, particles are pumped toward the direction of the applied magnetic field. \cite{Qi:2008aa}
Therefore, it can also be viewed as 
a 1D pump. Based on the Thouless formula for the 1D pump, \cite{Thouless:1983fk} the number of the pumped
particle is rederived, which is given by the first Chern number. Here, the first Chern number is computed 
using the wave function under the magnetic field. 
The equivalence of the two formula gives an interesting relation between two kinds of Chern numbers.
We present numerical calculation of the first Chern number, and show that this relation is valid as far as the 
mass gap is open. The relationship between the second Chern number associated with the Berry curvature
under zero magnetic field and the first Chern number associated with the Berry curvature under a finite magnetic field
is one of the main results.

In Sec. \ref{s:BEC}, we consider the system with boundaries perpendicular to the magnetic field.
Then it is possible to define the center of mass of the occupied particles along the pumped direction, 
which played a central role in the 
experimental observation of the 1D pump. \cite{Wang:2013fk_pump,Nakajima:2016aa,Lohse:2016aa}
We show that the bulk-edge correspondence recently established for the 1D pump \cite{Hatsugai:2016aa}
is also valid for the 3D pump.
Thus, from the behavior of the center of mass as the function of time, we can compute the number of the pumped 
particle, and hence the second Chern number.

Finally in Sec. \ref{s:Streda}, we give an alternative method of computing the second Chern number 
(and hence, the number of the pumped particles) using the generalized Streda formula.\cite{Fukui:2016aa}
This method is based on the four dimensional Hamiltonian in discrete time  whose spectral asymmetry 
gives the chiral anomaly for the lattice fermion. 
Since the generalized Streda formula is the only one method of numerically computing the second Chern number
for generic models, the success in reproducing the 
exact Chern number for the present model is of significance.
In Sec. \ref{s:sd}, we give summary and discussion including outlook.

\section{Dirac model}
\label{s:D}

The main topic of this paper is to calculate the U(1) current for the Dirac fermion on the lattice, including
external scalar and pseudscalar fields which serve as nonuniform mass terms.
Calculations on the lattice, however, is quite complicated, so that we first examine the Dirac fermion defined in 
continuous space-time, which may be of help in Sec. \ref{s:WD}.

Let us directly calculate the vector current for the model described by the 
Lagrangian density
\begin{alignat}1
{\cal L}=\bar\psi\left(i\gamma^\mu D_\mu-\sigma-i\gamma_5\pi\right)\psi,
\label{LagCon}
\end{alignat}
where $\mu=0,1$ with $\gamma^0=\sigma^1$, $\gamma^1=-i\sigma^2$, and $\gamma_5=\sigma^3$ 
for a $d=1+1$ model,
whereas
$\mu=0,\cdots,3$ for a $d=3+1$ model with $\gamma$ matrices defined by 
\begin{alignat}1
\gamma^0=\begin{pmatrix}0&\bm1\\\bm1&0\end{pmatrix}, 
\quad \bm\gamma=\begin{pmatrix}0&-\bm\sigma\\\bm\sigma&0\end{pmatrix},
\quad \gamma_5=\begin{pmatrix}\bm1&0\\0&-\bm1\end{pmatrix}.
\end{alignat}
These satisfy $\{\gamma^\mu,\gamma^\nu\}=2g^{\mu\nu}$, where $g^{\mu\nu}=\mbox{diag}(1,-1,-1,-1)$, and
$\gamma_5=i\gamma^0\gamma^1\gamma^2\gamma^3$.
The covariant derivative under a background electro-magnetic field is defined by $D_\mu=\partial_\mu-ieA_\mu(x)$.
Note $e<0$ for electrons.
We have introduced external scalar and pseudscalar fields denoted by $\sigma$ and $\pi$.
This model has been studied in Ref.\cite{Goldstone:1981aa} for the fractional fermion numbers on solitons and   
in Ref. \cite{Sekine:2016ab} for the CME and AHE for an antiferromagnetic
topological insulator with broken time reversal and parity symmetries.
In this paper, we assume that $\sigma$ and $\pi$ depend on the coordinates $x=(t,\bm x)$ 
through a single parameter
$\theta(x)$ such that  
$\sigma(x)=m\cos\theta(x)$ and 
$\pi(x)=m\sin\theta(x)$.
Then, we readily see that $\sigma+i\gamma_5\pi=me^{i\gamma_5\theta}$ in Eq. (\ref{LagCon}).
This model is invariant not only U(1) but also axial U(1) transformations,
$\psi\rightarrow e^{i\gamma_5 \alpha}\psi$, $\bar\psi\rightarrow \bar\psi e^{i\gamma_5 \alpha}$, 
and $\theta\rightarrow \theta-2\alpha$.
In what follows, we calculate the U(1) vector current 
in the $m\rightarrow\infty$ limit. 

The U(1) vector current is defined by
\begin{alignat}1
\langle j^{\mu}(x)\rangle&=\langle0|\bar\psi(x)\gamma^\mu\psi(x)|0\rangle
\nonumber\\
&=-\lim_{x'\rightarrow x}\tr \gamma^\mu\langle0|T\psi(x)\bar\psi(x')|0\rangle, 
\label{CurCon}
\end{alignat}
where the propagator is given by
\begin{alignat}1
\langle0|T\psi(x)\bar\psi(x')|0\rangle=\frac{i}{i\slashed D-me^{i\gamma_5\theta}+i\epsilon}\delta(x-x').
\end{alignat}
The positive infinitesimal constant $\epsilon$ will be sometimes suppressed for simplicity below.
Inserting 
\begin{alignat}1
\delta(x-x')=\int\frac{d^dk}{(2\pi)^d}e^{ik(x-x')},
\end{alignat}
with $kx=\omega t-\bm k\cdot\bm x$ and $d=2$ $(4)$ for a $1+1$ ($1+3$) dimensional system, 
we can write Eq. (\ref{CurCon}) as 
\begin{alignat}1
\langle j^\mu(x)\rangle=\int\frac{d^dk}{i(2\pi)^d}e^{-ikx}\tr\gamma^\mu
\frac{1}{i\slashed D-me^{i\gamma_5\theta}+i\epsilon}e^{ikx}.
\label{CurCon2}
\end{alignat}
To calculate the topological sector of the current, it is convenient to use the identity
\begin{alignat}1
&\frac{1}{i\slashed D-me^{i\gamma_5\theta}}=(-i\slashed D-me^{-i\gamma_5\theta})
\frac{1}{\slashed D^2+m^2+me^{i\gamma_5\theta}\gamma_5\slashed\partial\theta}.
\label{ConIde}
\end{alignat}
In the denominator, we further have
$\slashed D^2
=D^\mu D_\mu-\frac{i\gamma^\mu\gamma^\nu}{2}eF_{\mu\nu}$, 
where  $F_{\mu\nu}=\partial_\mu A_\nu-\partial_\nu A_\mu$ 
is the field strength of the background electro-magnetic field.
Note also $e^{-ikx}\slashed D e^{ikx}=\slashed D+i\slashed k$.
Finally, let us make a scale transformation for the momentum,
$k^\mu\rightarrow mk^\mu$, and expand the propagator with respect to $1/m$.
Then, Eq. (\ref{CurCon2}) becomes
\begin{widetext}
\begin{alignat}1
\langle j^\mu(x)\rangle
&=\int\frac{d^dk}{i(2\pi)^d}m^{d-2}\tr\gamma^\mu
(-i\slashed D+m\slashed k-me^{-i\gamma_5\theta})
\frac{1}{1-k^2+e^{i\gamma_5\theta}\gamma_5\slashed\partial\theta/m-
\frac{i\gamma^\rho\gamma^\sigma}{2} eF_{\rho\sigma}/m^2+\mathcal{O}/m^2}
\nonumber\\
&=\int\frac{d^dk}{i(2\pi)^d}\sum_{n=0}^\infty\frac{m^{d-1-n}}{(1-k^2-i\epsilon)^{n+1}}
\tr\gamma^\mu
\left(-i\frac{\slashed{D}}{m}+\slashed k-e^{-i\gamma_5\theta}\right)
\left(-e^{i\gamma_5\theta}\gamma_5\slashed\partial\theta+
\frac{i\gamma^\rho\gamma^\sigma}{2m}{eF}_{\rho\sigma}+\frac{\mathcal{O}}{m}\right)^n,
\label{CurCon3}
\end{alignat}
where $k^2=\omega^2-\bm k^2$ and $\mathcal{O}=2ik^\mu D_\mu+D^\mu D_\mu$.
The topological sector of the current is associated with the terms including $\gamma_5$.
For the $d=1+1$ system, using  (\ref{TraGam1D}),
it turns out that the $n=1$ term in Eq. (\ref{CurCon3}) survives in the $m\rightarrow\infty$ limit.
Thus, we have
\begin{alignat}1
\langle j^\mu(x)\rangle
&=\int\frac{d^2k}{i(2\pi)^2}\frac{\tr\gamma^\mu\gamma_5\gamma^\nu\partial_\nu\theta}{(1-k^2-i\epsilon)^2}
+O(m^{-1})
=\frac{1}{2\pi}\epsilon^{\mu\nu}\partial_\nu\theta,
\label{DCur1}
\end{alignat}
where we have used Eq. (\ref{Int_1}).
For the $d=3+1$ system, using (\ref{TraGam3D}),
we see that among terms with $\gamma_5$, only those included in the $n=2$ term in Eq. (\ref{CurCon3}) survive, 
\begin{alignat}1
\langle j^\mu(x)\rangle
&=\int\frac{d^4k}{i(2\pi)^4}\frac{m}{(1-k^2-i\epsilon)^{3}}
\tr\gamma^\mu\cdot 2
\gamma_5\gamma^\nu\partial_\nu\theta
\frac{i\gamma^\rho\gamma^\sigma}{2m}{eF}_{\rho\sigma}+O(m^{-1})
=-\frac{e}{8\pi^2}\epsilon^{\mu\nu\rho\sigma}(\partial_\nu\theta) F_{\rho\sigma}.
\label{DCur3_1}
\end{alignat}
\end{widetext}
This result can be written more explicitly as 
\begin{alignat}1
&\rho(x)=\frac{-e}{4\pi^2}\nabla\theta(x)\cdot\bm B(x),
\nonumber\\
&\bm j(x)=\frac{e}{4\pi^2}\left[\dot\theta(x)\bm B(x)+\nabla\theta(x)\times\bm E(x)\right].
\label{DCur3_2}
\end{alignat}
Thus, we have reached the known result.\cite{Wilczek:1987ab,Qi:2008aa}
In the next section, we use a similar method to calculate the current for the Wilson-Dirac model.

\section{Wilson-Dirac model}
\label{s:WD}

In this section, we calculate the U(1) current density for the Wilson-Dirac model
based on a similar technique demonstrated in Sec. \ref{s:D}.

\subsection{Lattice action}
\label{s:WD_la}
 
We consider the Wilson-Dirac Hamiltonian\cite{Qi:2008aa} defined on 
the 1D or 3D regular lattices, 
\begin{alignat}1
&H=\sum_{\bm x}a^{d-1}\psi^\dagger(t,\bm x)
{\cal H}(t,\bm x)
\psi(t,\bm x),
\nonumber\\
&{\cal H}(t,\bm x)\equiv -i\bm\alpha\cdot \bm D^{\rm L}
-\beta\left(\frac{m}{a} e^{i\gamma_5\theta}+\frac{ba}{2}\Delta^{\rm L}\right),
\label{WDHam}
\end{alignat}
where $a$ is the lattice constant, 
and $\bm\alpha$ and $\beta$ are given by the $\gamma$-matrices in Sec. \ref{s:D},
$\bm\alpha=\beta\bm\gamma$ and $\beta=\gamma^0$.
The mass is written by $m/a$, where $m$ is a dimensionless 
parameter. We will keep $m$ finite as $a\rightarrow0$. This corresponds to 
taking $m\rightarrow\infty$ limit in the continuum theory in Sec. \ref{s:D}.  
The lattice operators are introduced by 
\begin{alignat}1
D_j^{\rm L}&=\frac{1}{2}(\nabla_j+\nabla_j^*),\quad
\Delta^{\rm L}=\frac{1}{a}\sum_{j}(\nabla_j-\nabla_j^*),
\label{LatDirOpe}
\end{alignat}
where the forward and backward covariant differences are defined by
\begin{alignat}1
&\nabla_j\psi(t,\bm x)=\frac{1}{a}\left[U_j(t,\bm x)\psi(t,\bm x+a\hat j)-\psi(t,\bm x)\right],
\nonumber\\
&\nabla_j^*\psi(t,\bm x)=\frac{1}{a}\left[\psi(t,\bm x)-U^\dagger_j( t,\bm x-a\hat j)\psi(t,\bm x-a\hat j)\right],
\label{CovDifOpe}
\end{alignat}
with the gauge field $U_j(t,\bm x)=e^{-ieaA_j(t,\bm x)}$, and 
$\hat j$ stands for the unit vector in the $j$ direction.
The term associated with the Laplacian on the lattice, $\Delta^{\rm L}$, is 
the Wilson term originally introduced to avoid the doubling of fermions. 
It violates the axial U(1) symmetry and is the origin of the axial anomaly. 
In the condensed matter physics literature,
it is known to control the topological property of the ground state.
In this paper, we consider the case $b>0$.
We derive the current of the lattice model in the limit $a\rightarrow0$ in a way similar to the continuum model
in Sec. \ref{s:D}.

We treat time $t$ as continuous variable, so that the lattice action reads
\begin{alignat}1
S=\int_{-\infty}^\infty dt\sum_{\bm x}a^{d-1}\bar\psi(x)
\left(
i\slashed D^{\rm L}-\frac{m}{a}e^{i\gamma_5\theta}-\frac{ba}{2}\Delta^L
\right)\psi(x),
\label{LatAct}
\end{alignat}
where $x=(t,\bm x)$ and the covariant derivative with respect to $t$ is the same as that in the continuum model,
$D_0\equiv D_0^{\rm L}=\partial_0-ieA_0$.

The fermion propagator on the lattice is 
\begin{alignat}1
\langle0&|T\psi(x)\bar\psi(x')|0\rangle
\nonumber\\
&=\frac{1}{i\slashed D^{\rm L}-\frac{m}{a}e^{i\gamma_5\theta}-\frac{ba}{2}\Delta^{\rm L}+i\epsilon}\frac{i}{a^{d-1}}
\delta(t-t')\delta_{\bm x,\bm x'} ,
\label{WilDirPro}
\end{alignat}
where $\epsilon$ is a positive infinitesimal constant
to implement the  time ordering, which will be sometimes suppressed for simplicity below.
This follows from the identity
\begin{alignat}1
0&=\frac{1}{Z}\int{\cal D}\psi{\cal D\bar\psi}\frac{\delta}{\delta\bar\psi(x)}\left[e^{iS}\bar\psi(x')\right]
\nonumber\\
&=ia^{d-1}\left(i\slashed D^{\rm L}-\frac{m}{a}e^{i\gamma_5\theta}-\frac{ba}{2}\Delta^{\rm L}\right)
\langle\psi(x)\bar\psi(x')\rangle
\nonumber\\
&\qquad+\delta(t-t')\delta_{\bm x,\bm x'} .
\end{alignat}

\subsection{Conserved U(1) current}
\label{s:WD_cc}

The action (\ref{LatAct}) is invariant under the gauge transformation
\begin{alignat}1
&\psi'(x)=\Lambda(x)\psi(x),
\nonumber\\
&\bar\psi'(x)=\bar\psi(x)\Lambda^\dagger(x),
\nonumber\\
&eA_0'(x)=eA_0(x)-i\Lambda^\dagger(x)\partial_0\Lambda(x),
\nonumber\\
&U'_j(x)=\Lambda(x)U_j(x)\Lambda^\dagger(x+a\hat j).
\end{alignat} 
For the U(1) gauge theory, the conserved current density can be obtained by considering 
an infinitesimal gauge transformation,
$\Lambda(x)=1+i\lambda(x)$, which induces
\begin{alignat}1
&\delta A_0(x)=\partial_0\lambda(x),
\nonumber\\
&\delta U_j(x)=-ia\left(\partial_j\lambda(x)\right)U_j(x),
\nonumber\\
&\delta U_j^\dagger(x-a\hat j)=ia\left(\partial_j^*\lambda(x)\right)U_j^\dagger(x-a\hat j),
\label{InfGauTra}
\end{alignat}
where we have introduced the forward and backward differences by
\begin{alignat}1
&\partial_j\lambda(x)=\frac{1}{a}\left[\lambda(x+a\hat j)-\lambda(x)\right],
\nonumber\\
&\partial^*_j\lambda(x)=\frac{1}{a}\left[\lambda(x)-\lambda(x-a\hat j)\right].
\label{DifOpe}
\end{alignat}
It is only in this subsection \ref{s:WD_cc} to use the symbols $\partial_j$ and $\partial_j^*$ as differences. 
The change of the action (\ref{LatAct}) under the infinitesimal gauge transformation reads 
\begin{widetext}
\begin{alignat}1
\delta S&=\int dt\sum_{\bm x}a^{d-1}\Big\{
\bar\psi(x)i\gamma^0\left[-ie\delta A_0(x)\right]\psi(x)
+\frac{i}{2a}\bar\psi(x)\sum_j\gamma^j
\left[\delta U_j(x)\psi(x+a\hat j)-\delta U_j^\dagger(x-a\hat j)\psi(x-a\hat j)\right]
\nonumber\\
&\qquad\qquad
-\frac{b}{2a}\bar\psi(x)\sum_j
\left[\delta U_j(x)\psi(x+a\hat j)+\delta U_j^\dagger(x-a\hat j)\psi(x-a\hat j)\right]
\Big\} .
\end{alignat}
Substituting the infinitesimal gauge transformation (\ref{InfGauTra}) into the above equation and using the relations
$\sum_x\partial_jf(x)g(x)=-\sum_xf(x)\partial_j^* g(x)$,
and 
$\partial_jf(x)=\partial_j^*f(x+a\hat j)$,
we have
\begin{alignat}1
\delta S&=-\int dt\sum_xa^{d-1}\lambda(x)\Big\{
\partial_0\left[\bar\psi(x)\gamma^0\psi(x)\right]
+\frac{1}{2}\sum_j\partial_j^*\left[\bar\psi(x)\gamma^jU_j(x)\psi(x+a\hat j)
+\bar\psi(x+a\hat j)\gamma^jU_j^\dagger(x)\psi(x)\right]
\nonumber\\
&\qquad\qquad+\frac{ib}{2}\sum_j\partial_j^*\left[
\bar\psi(x) U_j(x)\psi(x+a\hat j)-\bar\psi(x+a\hat j)U_j^\dagger(x)\psi(x)\right]
\Big\}
\nonumber\\
&\equiv
-\int dt\sum_{\bm x}a^{d-1}\lambda(x)\left[\partial_0j^0(x)+\sum_l\partial_l^*j^l(x)\right] ,
\end{alignat}
where $\partial_0=\partial_t$ is the derivative with respect to $t$ and 
$\partial_j^*$ is the backward difference 
operator in Eq. (\ref{DifOpe}).
Thus, we reach the conserved U(1) current density 
\begin{alignat}1
&j^0(x)=\bar\psi(x)\gamma^0\psi(x),
\nonumber\\
&j^l(x)=\bar\psi(x)\gamma^l\psi(x)
+\frac{a}{2}
\left[
\bar\psi(x)\gamma^l\nabla_l\psi(x)+\bar\psi(x)\overleftarrow{\nabla}_l\gamma^l\psi(x)
\right]
+\frac{iba}{2}
\left[
\bar\psi(x)\nabla_l\psi(x)-\bar\psi(x)\overleftarrow{\nabla}_l\psi(x)
\right],
\label{WilDirCur}
\end{alignat}
\end{widetext}
where repeated $l$ in the middle term on the rhs of the second equation is not summed. 
We have also defined
\begin{alignat}1
a\bar\psi(x)\overleftarrow{\nabla}_l\equiv\bar\psi(x+a\hat l)U^\dagger(x)-\bar\psi(x).
\end{alignat}
While the charge density is the same as that of the continuum theory,
the current density includes lattice effects described by the differences.
In what follows, we calculate the charge density and current density, separately.

\subsection{Charge density in the continuum limit}
\label{s:ChaDen}

The computation of the charge density $j^0(x)$ is much simpler 
than that of the current density, since we have regarded $t$ as continuous variable. 
Therefore, let us first start with the charge density, 
\begin{alignat}1
\langle j^0(x)\rangle&=
\langle0|\bar\psi(x)\gamma^0\psi(x)|0\rangle
\nonumber\\
&=
\lim_{x'\rightarrow x}(-)\tr \langle0|T\gamma^0\psi(x)\bar\psi(x')|0\rangle .
\end{alignat}
This can be calculated in a way similar to the continuum model. Namely, substituting the propagator 
(\ref{WilDirPro}) and inserting the plane-wave representation of the $\delta$ function, we have 
\begin{widetext}
\begin{alignat}1
\langle j^0(x)\rangle
&=
\lim_{x'\rightarrow x}\tr\gamma^0
\frac{1}{i\slashed D^{\rm L}-\frac{m}{a}e^{i\gamma_5\theta}-\frac{b}{2}a\Delta^{\rm L}
+i\epsilon}\frac{-i}{a^{d-1}}\delta(t-t')\delta_{\bm x,\bm x'} 
\nonumber\\
&=\frac{1}{a^{d-1}}\int_{-\infty}^\infty\frac{d\omega}{2\pi i}\int_{-\pi}^\pi\frac{d^{d-1}k}{(2\pi)^{d-1}}
e^{-i(\omega t-\frac{\bm k\cdot \bm x}{a})}
\tr\gamma^0
\frac{1}{i\slashed D^{\rm L}-\frac{m}{a}e^{i\gamma_5\theta}-\frac{b}{2}a\Delta^{\rm L}+i\epsilon}
e^{i(\omega t-\frac{\bm k\cdot \bm x}{a})}
\nonumber\\
&=\frac{1}{a^{d-1}}
\int_{-\infty}^\infty\frac{d\omega}{2\pi i}
\int\frac{d^{d-1}k}{(2\pi)^{d-1}}
e^{-i\frac{kx}{a}}
\tr\gamma^0
\frac{1}{ia\slashed D^{\rm L}-me^{i\gamma_5\theta}-\frac{b}{2}a^2\Delta^{\rm L}+i\epsilon}
e^{i\frac{kx}{a}},
\label{J0_1}
\end{alignat}
\end{widetext}
where in the last line, we have rescaled $\omega\rightarrow \omega/a$, and
$kx$ stands for the abbreviation of $\omega t-\bm k\cdot \bm x$. 
We will carry out the above integral in the limit $a\rightarrow 0$,
implying the large mass 
limit $m/a\rightarrow \infty$ in the continuum model.
Notice that
\begin{alignat}1
&e^{-i\frac{kx}{a}} a\nabla_j e^{i\frac{kx}{a}}=e^{-ik_j}a\nabla_j+e^{-ik_j}-1,
\nonumber\\
&e^{-i\frac{kx}{a}} a\nabla_j^* e^{i\frac{kx}{a}}=e^{ik_j}a\nabla_j^*-e^{ik_j}+1,
\end{alignat}
where $j$ denotes the spatial direction.
Therefore, in the limit $a\rightarrow 0$, the difference becomes
\begin{alignat}1
e^{-i\frac{kx}{a}} aD^{\rm L}_j e^{i\frac{kx}{a}}
&=-i\sin k_j+\cos k_j aD_j +O(a^2)
\nonumber\\
&\equiv -is_j+a\widetilde D_j +O(a^2),
\label{DiraOpeCon_1}
\end{alignat}
where $D_j$ in the first line is the covariant derivative in the continuum limit in Sec. \ref{s:D},
and in the second line, the abbreviations $s_j$ (and $c_j$) mean
$s_j\equiv \sin k_j$ (and $c_j=\cos k_j$), 
and repeated $j$ in the definition of $\widetilde D_j=c_jD_j$ is not summed.
As for the time component, we simply have
\begin{alignat}1
e^{-i\frac{kx}{a}} aD^{\rm L}_0 e^{i\frac{kx}{a}}&=i\omega+aD_0
\equiv is_0+a\widetilde D_0.
\label{DiraOpeCon_2}
\end{alignat}
Thus, the differences $D_\mu^{\rm L}$ in Eqs. (\ref{DiraOpeCon_1})  and (\ref{DiraOpeCon_2})
are summarized as
\begin{alignat}1
e^{-i\frac{kx}{a}} aD^{\rm L}_\mu e^{i\frac{kx}{a}}&=is_\mu+a\widetilde D_\mu +O(a^2),
\label{aLDt}
\end{alignat}
where $s_\mu=(\omega,-s_j)$, and $\widetilde D_\mu=c_\mu D_\mu$ (no summation over $\mu$) 
with $c_\mu=(1,c_j)$.
Likewise, the Laplacian on the lattice becomes
\begin{alignat}1
e^{-i\frac{k\cdot x}{a}} a^2\Delta^{\rm L} e^{i\frac{k\cdot x}{a}}
&=2\sum_{j=1}^{d-1}\left(\cos k_j-1-i\sin k_j aD_j\right) +O(a^2)
\nonumber\\
&=2\left(c_{\rm s}-ia\widetilde D_{\rm s}\right)+O(a^2),
\label{LLapt}
\end{alignat}
where $c_{\rm s}=\sum_{j=1}^{d-1} (c_j-1)$ and $\widetilde D_{\rm s}\equiv\sum_{j=1}^{d-1} s_jD_j$.
Using these, the propagator in Eq. (\ref{J0_1}) can be written as
\begin{widetext}
\begin{alignat}1
e^{-i\frac{kx}{a}}&
\frac{1}
{ia\slashed D^L-me^{i\gamma_5\theta}-\frac{b}{2}a^2\Delta^L}
e^{i\frac{kx}{a}}
=\frac{1}{i(i\slashed{s}+a\slashed{\widetilde D})
-me^{i\gamma_5\theta}-b(c_{\rm s}-ia\widetilde D_{\rm s})}
\nonumber\\
&=\left\{-i(i\slashed{s}+a\widetilde{\slashed D})-me^{-i\gamma_5\theta}
-b(c_{\rm s}-ia\widetilde D_{\rm s})\right\}
\nonumber\\
&\qquad\qquad\times\frac{1}
{
\mu^2-s^2+me^{i\gamma_5\theta}\gamma_5a \tilde{\slashed{\partial}}\theta
-mbe^{-i\gamma_5\theta}\gamma_5a\tilde\partial_{\rm s}\theta
-\frac{i\gamma^\rho\gamma^\sigma}{2}ea^2\widetilde F_{\rho\sigma}
+ib\gamma^\rho ea^2\widetilde F_{\rho,{\rm s}}+\tilde{\mathcal{O}}
},
\label{LatProMom}
\end{alignat}
\end{widetext}
where $s^2=\omega^2-\bm s^2$, $\tilde{\partial}_\mu\equiv c_\mu\partial_\mu$ (no sum over $\mu$), 
$\tilde\partial_{\rm s}=\sum_{j=1}^{d-1} s_j\partial_j$, and
\begin{alignat}1
\mu^2=m^2\sin^2\theta+(m\cos\theta+bc_{\rm s})^2 .
\end{alignat}
We have also introduced two kinds of the field strength: 
First, $\widetilde F_{\mu\nu}\equiv c_\mu c_\nu F_{\mu\nu}$ (no sum over $\mu,\nu$) which follows from 
\begin{alignat}1
[\widetilde D_\mu,\widetilde D_\nu]=-ie\widetilde F_{\mu\nu},
\end{alignat}
and second, $\widetilde F_{\mu,{\rm s}}\equiv \sum_{j=1}^3c_\mu s_j F_{\mu j}$ (no sum over $\mu$) 
coming from 
\begin{alignat}1
[\widetilde D_\mu,\widetilde D_{\rm s}]=-ie\widetilde F_{\mu,{\rm s}},
\end{alignat}
where $F_{\mu\nu}$ is the field strength of the electro-magnetic field in the continuum model
defined in Sec. \ref{s:D}.
In Eq. (\ref{LatProMom}) the other operators without $\gamma$-matrices are simply denoted as 
$\tilde{\mathcal{O}}\equiv a^2\widetilde D^\mu\widetilde D_\mu+2is^\mu a\widetilde{D}_\mu
-b^2(a^2\widetilde{D}_{\rm s}^2+2ic_{\rm s}a\widetilde{D}_{\rm s})-2imb\cos\theta a\widetilde{D}_{\rm s}$. 
Furthermore, we have ignored the $O(a^2)$ terms in Eqs. (\ref{aLDt}) 
and (\ref{LLapt}) since they do not contribute to the charge density in the continuum 
limit. 

For sufficiently small $a$, we can expand the denominator on the rhs of  
Eq. (\ref{LatProMom}). The charge density (\ref{J0_1}) then can be written as 
\begin{widetext}
\begin{alignat}1
\langle j^0(x)\rangle
&=\frac{1}{a^{d-1}}
\int_{-\infty}^\infty\frac{d\omega}{2\pi i}
\int\frac{d^{d-1}k}{(2\pi)^{d-1}}\sum_{n=0}^\infty\frac{1}{(\mu^2-s^2-i\epsilon)^{n+1}}
\tr\gamma^0
\left(\slashed{s}-me^{-i\gamma_5\theta}-bc_{\rm s}-ia\widetilde{\slashed D}+ia\widetilde D_{\rm s}\right)
\nonumber\\
&\times
\left(
-me^{i\gamma_5\theta}\gamma_5a \tilde{\slashed{\partial}}\theta
+mbe^{-i\gamma_5\theta}\gamma_5a\tilde\partial_{\rm s}\theta
+\frac{i\gamma^\rho\gamma^\sigma}{2}ea^2\widetilde F_{\rho\sigma}
-ib\gamma^\rho ea^2\widetilde F_{\rho,{\rm s}}+\tilde{\mathcal{O}}
\right)^{n}.
\label{J0_2}
\end{alignat}
This equation for the lattice Wilson-Dirac fermion corresponds to Eq. (\ref{CurCon3}) for the continuum Dirac fermion.

\subsubsection{$d=1+1$ system}

We are interested in the terms with $\gamma_5$ in Eq. (\ref{J0_2}) which survive in the limit $a\rightarrow0$. 
Due to Eq. (\ref{TraGam1D}), it is enough to consider the $n=1$ term in Eq. (\ref{J0_2}).
\begin{alignat}1
\langle j^0(x)\rangle
&=\frac{1}{a}
\int_{-\infty}^\infty\frac{d\omega}{2\pi i}\int_{-\pi}^\pi\frac{dk_1}{2\pi}
\frac{
\tr\gamma^0\left(\slashed{s}-me^{-i\gamma_5\theta}-bc_{\rm s}\right)(-ma\gamma_5)
\left(e^{i\gamma_5\theta} \tilde{\slashed{\partial}}\theta
-be^{-i\gamma_5\theta}\tilde\partial_{\rm s}\theta
\right)
}
{(\mu^2-s^2-i\epsilon)^2} .
\label{ChaDen1Ori}
\end{alignat}
It turns out that the trace in the numerator of the above equation yields
\begin{alignat}1
(-ma)\left[
\tr\gamma_5\gamma^0\left(me^{-i\gamma_5\theta}+bc_{\rm s}\right)
e^{i\gamma_5\theta}\tilde{\slashed{\partial}}\theta
-b\tr\gamma_5\gamma^0\slashed se^{-i\gamma_5\theta}\tilde\partial_{\rm s}\theta)
\right] 
=2am\left[(m+bc_{\rm s}\cos\theta)c_1+b\cos\theta s_1^2\right]\epsilon^{01}\partial_1\theta .
\end{alignat}
\end{widetext}
The denominator becomes $\mu^2-s^2=\mu^2+\bm s^2-\omega^2\equiv \Omega^2(k_1,\theta)-\omega^2$, where
we have introduced a generic expression for later convenience,
\begin{alignat}1
\Omega^2(\bm k,\theta)&\equiv\bm s^2+\mu^2
\nonumber\\
&
=\sum_{j=1}^{d-1}s_j^2+m^2\sin^2\theta+\left[m\cos\theta+b\sum_{j=1}^{d-1}(c_j-1)\right]^2.
\label{OmeD}
\end{alignat}
Note $d=2$ in the present 1D system.
Then, using the integration over $\omega$ in Eq. (\ref{IntOme1}),
we finally obtain
\begin{alignat}1
\langle j^\mu(x)\rangle=G_1(\theta)\epsilon^{\mu\nu}\partial_\nu\theta,
\label{ChaDen1}
\end{alignat}
where we have derived the $\mu=0$ charge density in this subsection, although Eq. (\ref{ChaDen1}) is valid 
for the $\mu=1$ current density, as we will show in Appendix \ref{s:CurDen}, and we have introduced
\begin{alignat}1
&G_{d-1}(\theta)=N_{d-1}\int_{-\pi}^\pi d^{d-1}k
\frac{\Theta(\bm k,\theta)}
{\Omega^{d+1}(\bm k,\theta)},
\nonumber\\
&\Theta(\bm k,\theta)\equiv m\left[m+b\sum_{j=1}^{d-1}(\sec k_j-1)\cos\theta\right]\prod_{j=1}^{d-1}\cos k_j, 
\label{Gd}
\end{alignat}
with $N_1=1/(4\pi)$. It should be noted that $G_{d-1}(\theta)$ is independent of the electro-magnetic field,
and furthermore,  it depends on $x$ only through $\theta$. Thus, 
\begin{alignat}1
\int_0^{2\pi}G_1(\theta)d\theta=c_1
\label{FirChe}
\end{alignat}
does not depend on $x$ any longer. This is nothing but the first Chern number, which can also be computed 
using the wave functions of the Wilson-Dirac Hamiltonian (\ref{WDHam}).
See Sec. \ref{s:E}. 

In the case of $\theta=\theta(x)$,
Eq. (\ref{ChaDen1}) as well as Eq. (\ref{DCur1}) have attracted much interest as the topological number on solitons in 
quantum field theory.\cite{Goldstone:1981aa,Qi:2008aa}
On the other hand, when $\theta=\theta(t)$, we nowadays 
know that they describe the topological pump, which is of current interest.
Both phenomena mentioned above are related with the anomaly in two dimensional Dirac fermions, as will be discussed in Sec. \ref{s:sd}.

\subsubsection{$d=3+1$ system}

We are also interested in the terms with $\gamma_5$ in Eq. (\ref{J0_2}) which survive in the limit $a\rightarrow0$. 
Considering Eq. (\ref{TraGam3D}), it is enough to take the $n=2$ term in Eq. (\ref{J0_2}) into account.
\begin{widetext}
\begin{alignat}1
\langle j^0(x)\rangle
&=\frac{1}{a^3}
\int_{-\infty}^\infty\frac{d\omega}{2\pi i}
\int\frac{d^3k}{(2\pi)^3}
\frac{1}{(\mu^2-s^2-i\epsilon)^3}
\nonumber\\
&\times
\tr\gamma^0\left(\slashed{s}-me^{-i\gamma_5\theta}-bc_{\rm s}\right)
\left(
-me^{i\gamma_5\theta}\gamma_5a \tilde{\slashed{\partial}}\theta
+mbe^{-i\gamma_5\theta}\gamma_5a\tilde\partial_{\rm s}\theta
+\frac{i}{2}\gamma^\rho\gamma^\sigma ea^2\widetilde F_{\rho\sigma}
-ib\gamma^\rho ea^2\widetilde F_{\rho,{\rm s}}
\right)^2 .
\label{J0_3}
\end{alignat}
Among various terms in the above equation associated with the trace of the $\gamma$ matrices, 
the product terms between $\partial\theta$ and $\widetilde F$ 
give finite contributions, which have indeed $a^3$, 
\begin{alignat}1
ea^3\Big\{&
\tr\gamma^0(-me^{-i\gamma_5\theta}-bc_{\rm s})(-me^{i\gamma_5\theta}\gamma_5a\tilde{\slashed\partial}\theta)
(i\gamma^\rho\gamma^\sigma \widetilde F_{\rho\sigma})
+\tr\gamma^0\slashed smbe^{-i\gamma_5\theta}\gamma_5\tilde\partial_{\rm s}\theta
(i\gamma^\rho\gamma^\sigma \widetilde F_{\rho\sigma})
\nonumber\\
&+2\tr\gamma^0\slashed s(-me^{i\gamma_5\theta}\gamma_5\tilde{\slashed\partial}\theta)
(-ib\gamma^\rho\widetilde F_{\rho,{\rm s}})\Big\}
\nonumber\\
&=-mea^3\Big\{
i(m+bc_{\rm s}\cos\theta)\tr\gamma_5\gamma^0
(\tilde{\slashed\partial}\theta)
\gamma^\rho\gamma^\sigma \widetilde F_{\rho\sigma}
-ib\cos\theta\tr\gamma_5\gamma^0\slashed s\tilde\partial_{\rm s}\theta
\gamma^\rho\gamma^\sigma \widetilde F_{\rho\sigma}
-2ib\cos\theta\tr\gamma_5\gamma^0\slashed s\tilde{\slashed\partial}\theta
\gamma^\rho\widetilde F_{\rho,{\rm s}}\Big\}
\nonumber\\
&=-4ea^3m\left[m+b\sum_{j=1}^3(\sec k_j-1)\cos\theta\right]\left(\prod_{j=1}^3\cos k_j\right)
\epsilon^{0\nu\rho\sigma}(\partial_\nu\theta) F_{\rho\sigma}.
\end{alignat}
\end{widetext}
Thus, we finally reach
\begin{alignat}1
\langle j^\mu(x)\rangle=-e\frac{G_3(\theta)}{4\pi}\epsilon^{\mu\nu\rho\sigma}\partial_\nu\theta(x) F_{\rho\sigma}(x), 
\label{ChaDen3}
\end{alignat}
where we have derived the $\mu=0$ charge density in this subsection, although Eq. (\ref{ChaDen3}) is valid 
for the $\mu=1,2,3$ current density, as we will show in Appendix \ref{s:CurDen}.
$G_3(\theta)$ is defined by Eq. (\ref{Gd}) with $N_3=3/(8\pi^2)$, which follows from  Eq. (\ref{IntOme2}). 
Note that integration of $G_3(\theta)$ over $\theta$ gives the second Chern number
\begin{alignat}1
\int_0^{2\pi}G_3(\theta)d\theta=c_2.
\label{SecChe}
\end{alignat}
It should be stressed here that $G_3(\theta)$ does not depends on the electro-magnetic field.
The limit $a\rightarrow 0$ implies, therefore, small field limit as well.
The Chern number $c_2$ is alternatively calculated using the Berry curvature
with respect to the eigenfunctions of the Wilson-Dirac Hamiltonian with zero fields.
See Ref. \cite{Qi:2008aa} or \cite{Fukui:2016aa}, and also 
Sec. \ref{s:E}. 

Equation (\ref{ChaDen3}) or its continuum version  (\ref{DCur3_1}) show the topological magneto-electric effects,
including CME and AHE. For Weyl semimetals, $\theta$ is induced by the Zeeman term violating time reversal symmetry and/or
energy imbalance between the Weyl nodes breaking the inversion symmetry. In contrast, for the present Wilson-Dirac fermion,
$\theta$ describes a rotation between two kinds of mass terms which keeps a finite mass gap in the spectrum.
Regardless of whether the system is massless or massive, 
the magneto-electric effects are associated with the chiral anomaly, as will be discussed 
in Sec. \ref{s:sd}.

\begin{table}[htb]
\begin{tabular}{cccc}
\hline\hline
$k_1$ &$k_2$ &$\qquad \xi_3\Omega\qquad$&$\Theta(k_1,\theta)$\\
\hline\hline
$0$& $0$ & $m$ &$m^2$\\
$\pi$& $0$ & $m-2b$ &$-m(m-2b)$\\
$0$& $\pi$ & $-m$ &$m^2$\\
$\pi$& $\pi$ & $-m-2b$ &$-m(m+2b)$\\
\hline
\end{tabular}
\caption{
List of $\xi_3\Omega=m\cos k_2+b(\cos k_1-1)$ and $\Theta(k_1,\theta)=m[m+b(\sec k_1-1)\cos k_2]\cos k_1$ 
at $k_\mu=0$ or $\pi$.
}
\label{t:1}
\end{table}

\subsection{Chern numbers}
\label{s:E}

For the study of the temporal pump in the next section \ref{s:MEP}, 
we need an explicit value of the Chern number. 
Therefore,  we calculate the first and second Chern numbers in Eqs. (\ref{FirChe}) and (\ref{SecChe}).
Fortunately, the Wilson-Dirac model is so simple that one can calculate the second Chern number exactly.

\subsubsection{First Chern number}

We can regard $k_1$ and $k_2\equiv\theta$ as the coordinates 
of a two-dimensional torus T$^2$. A mapping $f$ from T$^2$ to 
S$^2$ can be defined by
\begin{alignat}1
&\xi_1=\frac{\sin k_1}{\Omega(k_1,\theta)},\quad \xi_2=\frac{m\sin\theta}{\Omega(k_1,\theta)},
\nonumber\\
&\xi_3=\frac{m\cos\theta+b(\cos k_1-1)}{\Omega(k_1,\theta)},
\end{alignat}
where $\Omega(k_1,\theta)$ is given by Eq. (\ref{OmeD}) for $d=2$. 
It is known that the Chern number is the degree of the mapping which can be computed by
\begin{alignat}1
\mbox{deg }f=\frac{1}{\mbox{Vol(S}^2)}\int\frac{1}{2!}\epsilon^{\alpha\beta\gamma}\xi_\alpha
d\xi_\beta d\xi_\gamma,
\label{Degf1}
\end{alignat}
where $\mbox{Vol(S}^2)=4\pi$ and $d\xi_\alpha$ is the coordinate differential 1-form. Indeed, 
it is not difficult to rewrite Eq. (\ref{Degf1}) by $k_\mu$ ($\mu=1,2$) to show 
$\mbox{deg\:}f=c_1$, where $c_1$ is given by the lhs of Eq. (\ref{FirChe}) with $G_1(\theta)$ 
in Eq. (\ref{Gd}).

It is readily seen that only the points $k_\mu=0$ or $\pi$ on T$^2$ are mapped 
to $\xi_\pm=(0,0,\pm1)$ on S$^2$. The degree of the mapping $f$ is then given by
\begin{alignat}1
\mbox{deg }f=&\sum_{k\in f^{-1}(\xi_+)}
\mbox{sgn}\left[\Theta(k_1,\theta)\right],
\end{alignat}
where $\Theta(k_1,\theta)$ is defined in Eq. (\ref{Gd}) for $d=2$.
From Table \ref{t:1}, we can read the coordinates $k_\mu$ mapped to $\xi_+$ with its degree of mapping
(or winding number) $\pm1$.
For example, when $0<m<2b$, only $(0,0)$ is mapped to $\xi_+$ with a positive winding number $m^2>0$,
whereas others $(0,\pi)$, $(\pi,0)$, and $(\pi,\pi)$ are mapped to $\xi_-$. 
Thus, we find $c_1=1$ in this case. The other cases are likewise.
Thus, Table \ref{t:1} leads to 
\begin{alignat}1
\mbox{deg }f=c_1=
\left\{
\begin{array}{rl}
+1&(0<|m|<2b)\\
0&(2b<|m|)
\end{array}
\right. .
\end{alignat}

\begin{table}[htb]
\begin{tabular}{cccc}
\hline\hline
($k_1$, $k_2$, $k_3$) &$k_4$ &$\qquad \xi_5\Omega\qquad$
&$\Theta(\bm k,\theta)$\\
\hline\hline
no $\pi$& $0$ & $m$ &$m^2$\\
one $\pi$& $0$ & $m-2b$ &$-m(m-2b)$\\
two $\pi$& $0$ & $m-4b$ &$m(m-4b)$\\
three $\pi$& $0$ & $m-6b$ &$-m(m-6b)$\\
no $\pi$& $\pi$ & $-m$ &$m^2$\\
one $\pi$& $\pi$ & $-m-2b$ &$-m(m+2b)$\\
two $\pi$& $\pi$ & $-m-4b$ &$m(m+4b)$\\
three $\pi$& $\pi$ & $-m-6b$ &$-m(m+6b)$\\
\hline
\end{tabular}
\caption{
List of $\xi_5\Omega=m\cos k_4+b\sum_j^3(\cos k_j-1)$ and $\Theta(\bm k,\theta)=m[m+b\sum_j^3(\sec k_j-1)\cos k_4]\prod_j^3\cos k_j$ at
$k_\mu=0$ or $\pi$. ``no $\pi$'' in the first column means all $k_\mu=0$, whereas ``one $\pi$" means that 
one of $k_\mu=\pi$ and others are $0$, and so on.
}
\label{t:2}
\end{table}

\subsubsection{Second Chern number}

The second Chern number can be obtained by 
extending the computation of the first Chern 
number for the S$^2$ to S$^4$. 
Let us introduce $\xi_\mu$ ($\mu=1,\cdots,5$) by 
\begin{alignat}1
&\xi_j=\frac{\sin k_j}{\Omega(\bm k,\theta)},\quad (j=1,2,3),\quad \xi_4=\frac{m\sin\theta}{\Omega(\bm k,\theta)},
\nonumber\\
&\xi_5=\frac{m\cos\theta+b\sum_{l=1}^{3}(\cos k_l-1)}{\Omega(\bm k,\theta)},
\end{alignat}
where $\Omega(\bm k,\theta)$ is defined in Eq. (\ref{OmeD}) for $d=4$. 
These define a mapping $f$ from T$^4$  spanned by $(k_1,k_2,k_3,\theta\equiv k_4)$ to S$^4$ spanned by $\xi_\mu$ satisfying
$\xi_\mu^2=1$. The Chern number is the degree of the mapping 
given by 
\begin{alignat}1
\mbox{deg }f=\frac{1}{\mbox{Vol(S}^4)}\int\frac{1}{4!}\epsilon^{\alpha\beta\gamma\delta\epsilon}\xi_\alpha
d\xi_\beta d\xi_\gamma d\xi_\delta d\xi_\epsilon,
\label{Degf}
\end{alignat}
where Vol(S$^4$)$=8\pi^2/3$. It is straightforward to rewrite Eq. (\ref{Degf}) 
by $k_j$ and $\theta$ to show $\mbox{deg }f=c_2 $, where $c_2$ is given by the lhs of Eq. (\ref{SecChe}) 
with $G_3(\theta)$ in Eq. (\ref{Gd}).
It is readily seen that the points $k_\mu=0$ or $\pi$ on T$^4$ are mapped to $\xi_\pm=(0,0,0,0,\pm1)$ on S$^4$.
The degree of the mapping $f$ is then given by
\begin{alignat}1
\mbox{deg }f=&\sum_{k\in f^{-1}(\xi_+)}
\mbox{sgn}\left[\Theta(\bm k,\theta)\right] ,
\end{alignat}
where $\Theta(\bm k,\theta)$ is defined in Eq. (\ref{Gd}) for $d=4$.
Thus, the second Chern number can be obtained from Table \ref{t:2} in the similar way to the first Chern number: 
\begin{alignat}1
\mbox{deg }f=c_2=
\left\{
\begin{array}{rl}
+1&(0<|m|<2b)\\
-2\quad&(2b<|m|<4b)\\
+1&(4b<|m|<6b)\\
0&(6b<|m|)\\
\end{array}
\right. .
\label{ExpCheNum2}
\end{alignat}

\section{Magneto-electric pump}
\label{s:MEP}

In Secs. \ref{s:ChaDen} and \ref{s:CurDen}, we have established the U(1) current 
(\ref{ChaDen3}) in the 3D Wilson-Dirac model.
This result is obtained in the limit $a\rightarrow0$. It also implies that the result is valid 
only in a weak field limit. On the other hand, the pump is topological so that the result may be valid 
as long as the mass gap is open. 
To check this point, we study the pump by numerical calculations.
In this section, we restrict our discussions to the 3D temporal particle pump.

\subsection{3D pump}
\label{s:3dpump}

Assume that the electro-magnetic field is static, 
and consider the case in which $\theta$ depends only on $t$, $\theta=\theta(t)$ with a period $T$, 
$\theta(t+T)=\theta(t)$.
When $1/T\ll m$, we can regard the process of changing $t$ as an adiabatic process.
Then, integration of Eq. (\ref{ChaDen3}) over $t$ in one period yields the {\it pumped} particle density 
\begin{alignat}1
\bm q(x)=e\frac{c_2}{2\pi}\bm B(x),
\end{alignat}
where $c_2$ is the second Chern number defined in Eq. (\ref{SecChe}) and explicitly given by (\ref{ExpCheNum2}). 
We have stressed there that
the second Chern number (\ref{SecChe}) is that of the Wilson-Dirac model in zero field limit.
Namely, the Chern number can be computed directly using the eigenfunctions of the Hamiltonian without 
the magnetic field.

Without loss of generality, we assume the magnetic field in the $z$ direction, $(0,0,B)$.
The total pumped particle number $Q^z$ can be defined by integrating over the $xy$ surface,
\begin{alignat}1
Q^z=ec_2\frac{\Phi}{2\pi},
\label{Qz1}
\end{alignat}
where $\Phi$ is the total flux penetrating the surface of an area under consideration.
This integration can be simply carried out, since $c_2$ is independent of $B$, as stressed already.
For numerical computations, we further consider a simpler case where the magnetic field
$B$ is uniform, and system is periodic in $xy$ surface whose size in the $x$ ($y$) direction is $N_x$ ($N_y$). 
Then, $\Phi=B a^2N_xN_y$, and the flux per plaquette should be rational
\begin{alignat}1
|e|\phi=|e|Ba^2=\frac{2\pi p}{q}\equiv \phi_0,
\label{FluPla}
\end{alignat}
where $p$ and $q (>0)$ are integers describing  a rational magnetic flux per plaquette.
\cite{Hofstadter:1976aa,Thouless:1982uq,kohmoto:85}
Thus, it turns out that the total pumped particle number $Q^z$ is given by
\begin{alignat}1
Q^z=\mbox{sgn(}e) c_2\frac{Ba^2N_xN_y}{2\pi}=\mbox{sgn(}e)c_2\frac{p}{q} N_xN_y.
\label{TotPumCha2}
\end{alignat}
The elementary pump is due to a nonzero second Chern number, and 
$pN_xN_y/q$ can be considered as the geometrical multiplicity of pumped particles associated with the
magnetic flux.

\subsection{3D pump as a set  of 1D pump}
\label{s:1dpump}

The magneto-electric pump discussed so far has deep relationship with the chiral anomaly, so that 
the $3+1$ dimensionality plays a crucial role.
However, the pumping itself is toward one direction of the applied magnetic field, and hence,
it can also be viewed as a 1D pump.
In this section, we derive the pumped particle number by the Thouless formula
for the 1D pump.\cite{Thouless:1983fk,Xiao:2010fk}

Consider the snapshot (for a fixed $t$, i.e., fixed $\theta$) many-body ground state eigenfunction obeying 
$H\ket{\Psi_n}=E_n\ket{\Psi_n}$. We assume that $\ket{\Psi_0}$ is the ground state and have a spectral gap 
$E_n(t)-E_0(t)>0$ at any time during the pump.
Then, the ground state $\ket{G}$ which satisfies the time-dependent Schr\"odinger equation at the first order 
of $\hbar$ is given by
\begin{alignat}1
\ket{G}=e^{-\frac{i}{\hbar}\int^tE_0 dt}\left(\ket{\Psi_0}
+i\hbar\sum_{n\ne0}\frac{\ket{\Psi_n}\braket{\Psi_n}{\dot\Psi_0}}{E_n-E_0}\right).
\label{PerG}
\end{alignat}
For non-interacting case,  it is enough to consider the 
single particle states.
Let ${\cal H}(\theta,\bm k;\bm B)$ be the Fourier-transformed  Hamiltonian given by Eq. (\ref{WDHam}),
and let $\psi(\theta,\bm k;\bm B)$ be the ground state (i.e., negative energy)  
multiplet wave functions of single-particle states satisfying
\begin{alignat}1
{\cal H}(\theta,\bm k;\bm B)\psi(\theta,\bm k;\bm B)=\psi(\theta,\bm k;\bm B){\cal E}(\theta,\bm k;\bm B),
\end{alignat}
where ${\cal E}(\theta,\bm k,\bm B)$ is the diagonal matrix of the energy eigenvalues.

Let us now consider the case where uniform magnetic field is applied in the $z$ direction, as studied in Sec. \ref{s:3dpump}.
Regarding $k_x$ and $k_y$ as parameters,  
the current toward the $z$ direction, $J^z$,
with respect to the state Eq. (\ref{PerG}) is given by \cite{Thouless:1983fk}
\begin{alignat}1
\langle J^z\rangle=
\int_0^{2\pi}\frac{dk_z}{2\pi i}f(\theta,k_z;k_x,k_y,B),
\end{alignat}
where $f(\theta,k_z;k_x,k_y,B)\equiv
\partial_{k_z}\psi^\dagger\partial_{\theta}\psi-\partial_\theta\psi^\dagger\partial_{k_z}\psi$
is the Berry curvature with respect to $\theta$ and $k_z$ with fixed $k_x$ and $k_y$.
Integrating $\langle J^z\rangle$ with respect to $t$ in one period as well as with respect to $k_x$ and $k_y$, 
we obtain the total number of pumped particle,
\begin{alignat}1
Q^z=\sum_{k_x,k_y}c_1(B),
\label{1ChaNum}
\end{alignat}
where 
\begin{alignat}1
c_1(B)=\frac{1}{2\pi i}\int_0^{2\pi}d\theta \int_0^{2\pi}dk_zf(\theta,k_z;k_x,k_y,B),
\label{SecFirCheNum}
\end{alignat}
is the first Chern number on the section specified by fixed $k_x,k_y$.
However, it should be noted that the Chern number $c_1(B)$ does depend on the applied magnetic field, but does not
depend on $k_x$ ($\tilde k_x$) and $k_y$, provided that the mass gap always opens.

When we compute the eigenfunctions in the Landau gauge 
$U_j(t,\bm x)=e^{-ieaA_j(t,\bm x)}$ with
\begin{alignat}1
e\bm A(\bm x)=(0,\mbox{sgn(}e) \phi\frac{x}{a},0),
\label{LanGau}
\end{alignat}
we can take the $q$ sites in the $x$-direction as a unit cell, 
and therefore, we set $N_x=q\tilde N_x$ for the periodic boundary condition. 
In this case, $k_y=2\pi n_y/N_y$ with $n_y=1,\cdots,N_y$ ,
whereas  $\tilde k_x=2\pi n_x/\widetilde N_x$ with $n_x=1,\cdots,\widetilde N_x$.
It follows from Eq. (\ref{1ChaNum}) and from the fact that $c_1(B)$ is independent of $k_x$ and $k_y$ that
\begin{alignat}1
Q^z=c_1(B,\mbox{sgn}(e))\tilde N_xN_y,
\label{Qz2}
\end{alignat}
where in Eq. (\ref{1ChaNum}) $k_x$ is replaced by $\tilde k_x$, and  
the dependence of $c_1$ on the sign of the charge $e$ through Eq. (\ref{LanGau}) has been explicitly denoted.
Comparing Eq. (\ref{TotPumCha2}), we have a simple relationship between the second Chern number $c_2$ in the 
zero magnetic field and $c_1$ in the magnetic field $B$,
\begin{alignat}1
c_1(B,\mbox{sgn}(e))=\mbox{sgn}(e) c_2 p,
\label{RelC1C2}
\end{alignat}
where $p$ is given by Eq. (\ref{FluPla}).
This relation may be useful for computing the second Chern number, since 
the numerical method of computing the first Chern number has already been established.\cite{FHS05}
 
 \begin{figure}[thb]
\begin{center}
\begin{tabular}{c}
\includegraphics[scale=0.8]{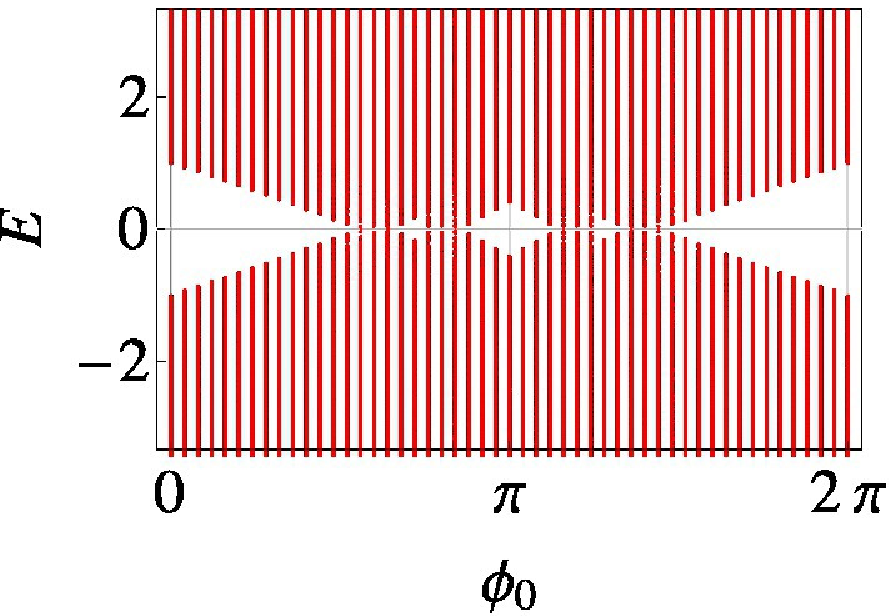}\\
\includegraphics[scale=0.8]{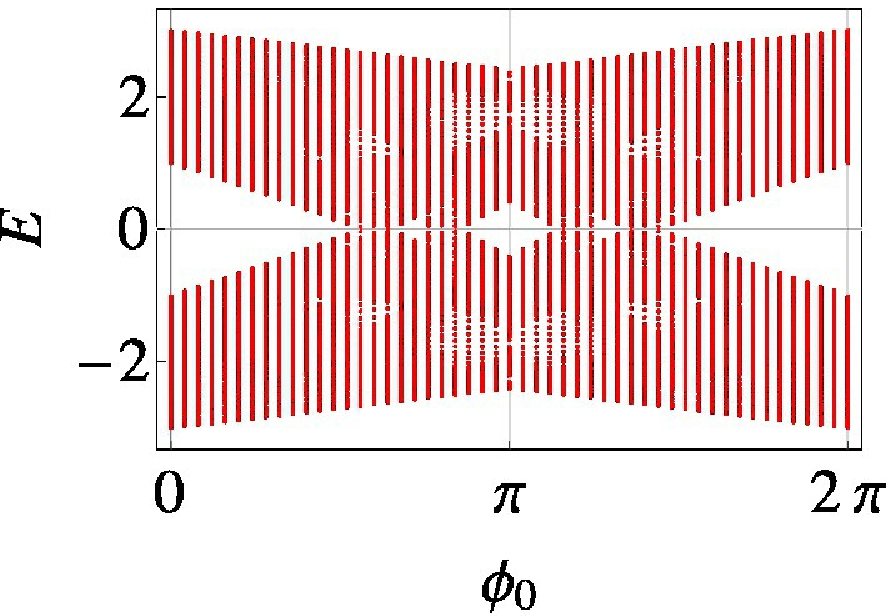}
\end{tabular}
\caption{
The Hofstadter butterfly diagrams for fixed $\theta=0$ and for $m=b$ (upper) and $m=3b$ (lower).
We set $q=50$, and the system size is $\widetilde N_x=1$ and $N_y=N_z=50(=q)$.
}
\label{f:but}
\end{center}
\end{figure}

\begin{table}[htb]
\begin{tabular}{cc|cc|cc||cc|cc|cc}
\hline
\multicolumn{6}{c||}{$m=b$}&\multicolumn{6}{c}{$m=3b$}\\
\hline\hline
$\frac{p}{q}$ 
& $c_1$&$\frac{p}{q}$ & $c_1$&$\frac{p}{q}$ & $c_1$&$\frac{p}{q}$  & $c_1$&$\frac{p}{q}$  & $c_1$&$\frac{p}{q}$  & $c_1$\\
\hline\hline
$\frac{1}{20}$&1 &&&&&$\frac{1}{20}$&$-2$ &&&&\\ 
$\frac{2}{20}$&2 &$\frac{1}{10}$&1&&&$\frac{2}{20}$&$-4$ &$\frac{1}{10}$&$-2$&&\\
$\frac{3}{20}$&3 &&&&&$\frac{3}{20}$&$-6$ &&&&\\
$\frac{4}{20}$&4 &$\frac{2}{10}$&2&$\frac{1}{5}$&1&$\frac{4}{20}$&$-8$ &$\frac{2}{10}$&$-4$&$\frac{1}{5}$&$-2$\\
\hline
\end{tabular}
\caption{
The (section) first Chern number (\ref{SecFirCheNum}) computed on the discretized $(\theta,k_z)$ 
Brillouin zone \cite{FHS05} in the case $e<0$ for several $\frac{p}{q}=\frac{\phi_0}{2\pi}$.
}
\label{t:3}
\end{table}

We show 
in Table \ref{t:3} 
the list of the section Chern number $c_1(B)$ in (\ref{SecFirCheNum}) for various $\phi_0$, and 
in Fig. \ref{f:but} the corresponding Hofstadter butterfly diagrams.\cite{Hofstadter:1976aa}
The Hofstadter diagrams tell that the mass gap at $\phi_0=0$ becomes smaller as a function of $\phi_0$,
and eventually closed around $\phi_0\sim \pi/2$ for both cases $m=b$ and $m=3b$.
Thus, it turns out from the Table \ref{t:3} 
that the relationship (\ref{RelC1C2}) is indeed valid as long as the mass gap is finite.
This implies the absence of the higher order corrections for the chiral anomaly. \cite{Adler:1969ab} 

\begin{figure*}[htb]
\begin{center}
\begin{tabular}{cc}
\hspace*{3mm}\includegraphics[scale=0.76]{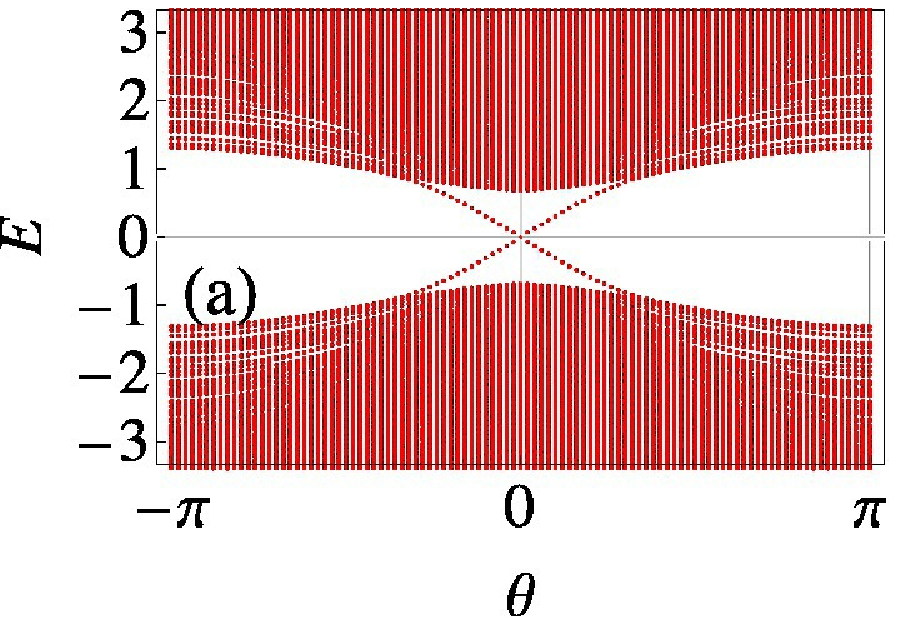}
&\hspace*{3mm}\includegraphics[scale=0.76]{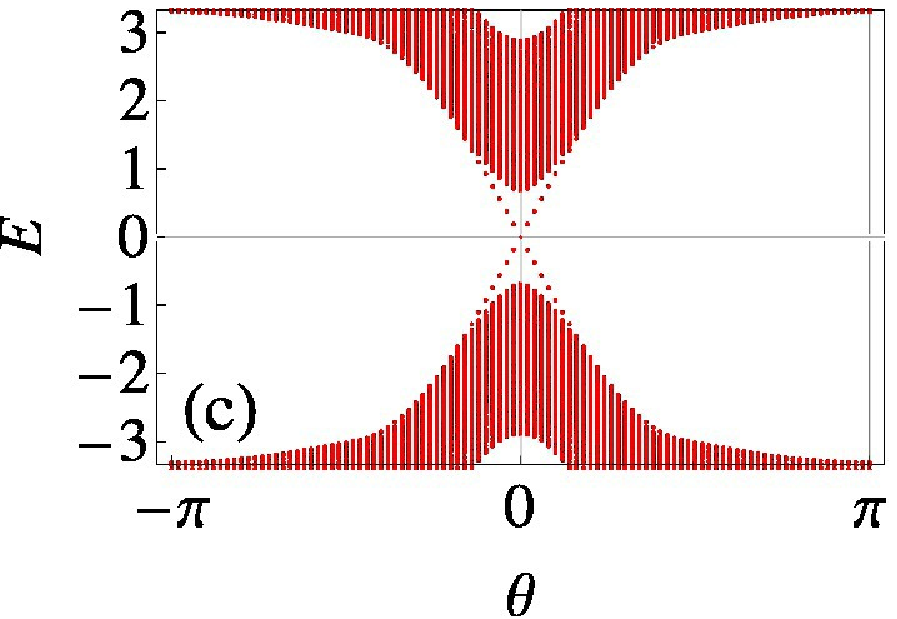}\\
\includegraphics[scale=0.8]{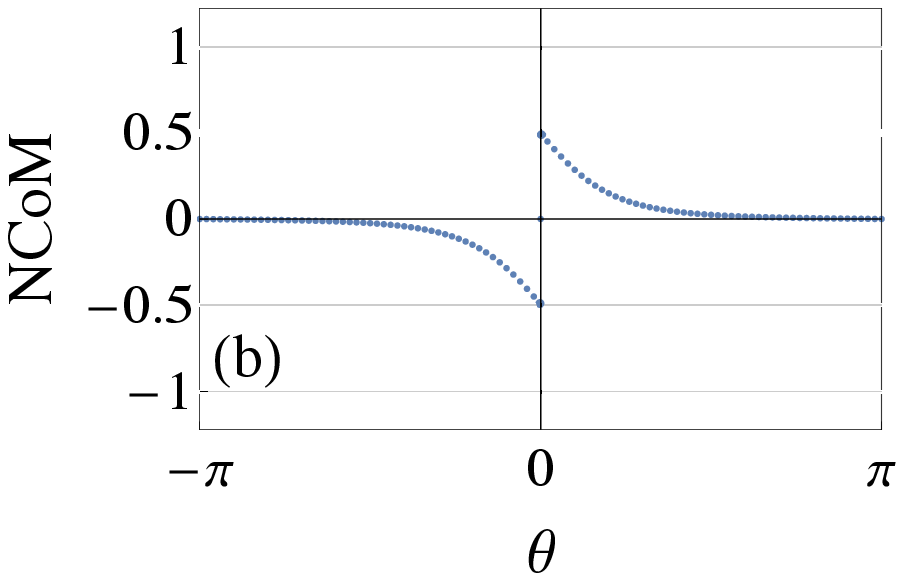}
&\includegraphics[scale=0.8]{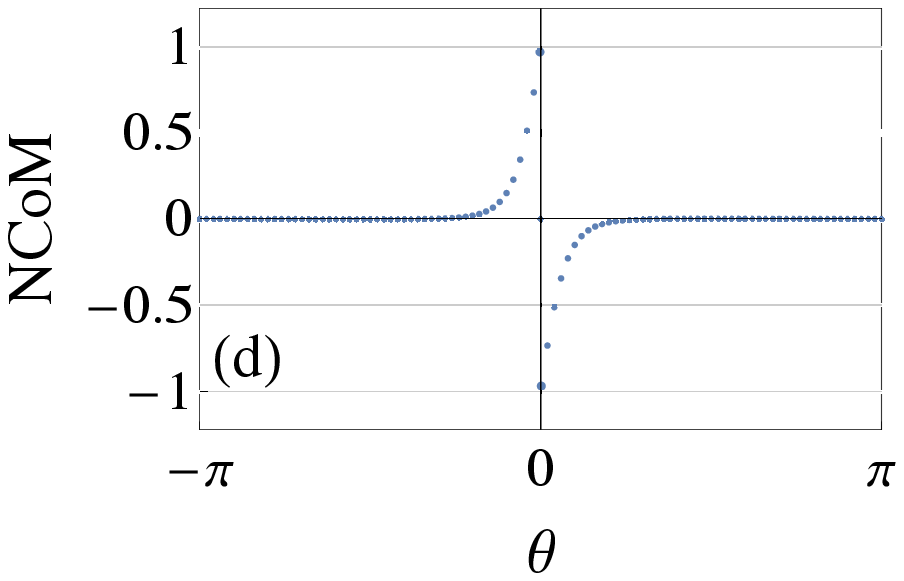}\\
\end{tabular}
\caption{
Upper two figures show 
the spectra as the function of $\theta$
of the Wilson-Dirac model for electrons ($e<0$) with top and bottom surfaces of width $N_z=20$ 
under $p/q=1/10$ magnetic flux per plaquette.
Lower two figures show 
the normalized center of mass defined by $p_z(\theta)$.
(a) and (b) are for the mass $m=b$ and (c) and (d) are for the mass $m=3b$.
The discontinuities are (b) $p_z(+0)-p_z(-0)=0.981$ and (d) $-1.938$, which suggest that
$c_2=1$ and $-2$. 
}
\label{f:com}
\end{center}
\end{figure*}

\subsection{Bulk-edge correspondence}
\label{s:BEC}

So far we have discussed the topological property of the bulk system. 
In this subsection, we discussed the surface states of the 3D topological pump, considering a system with boundaries.
In Ref.\cite{Hatsugai:2016aa}, the bulk-edge correspondence in the 1D Thouless pump has been discussed.
Consider the system with boundaries. Then one can define the center of mass of the occupied particles.
\cite{Wang:2013fk_pump}
It has been shown that the change of the center of mass is just the number of pumped particle.
\cite{Wang:2013fk_pump}
Consider here a system
coupled with a particle reservoir. Then,  the center of mass as a function of time
shows discontinuity due to sudden change of the ground state when the chemical potential crosses 
the edge states.\cite{Hatsugai:2016aa}
After one period, the center of mass returns to the initial value.
This implies that the amount of pumped particles in the bulk are compensated by these discontinuities.
Thus, from the discontinuities, one can know the number of the pumped particle.\cite{Hatsugai:2016aa}

Consider the Wilson-Dirac Hamiltonian with bottom and top surfaces,
labeled by $j_z=0$ and $j_z=N_z$, respectively, perpendicular to the $z$ axis.
Let $\psi^{n}(\theta,\tilde k_x,k_y;B)$ be the $n$th normalized eigenfunction of the Hamiltonian with the boundaries 
${\cal H}(\theta,\tilde k_x,k_y;B)$,
where we have assumed the Landau gauge in Sec. \ref{s:1dpump}.
Then, we can define the normalized center of mass of the ground state along the $z$ axis as
\begin{alignat}1
&p_z(\theta)=\frac{1}{p\widetilde N_x N_y}
\sum_{n~{\rm occ.}}\sum_{\tilde k_x,k_y}\sum_{j_z=0}^{N_z}\left(\frac{j_z}{N_z}-\frac{1}{2}\right)|\psi_{j_z}^{n}(\theta,\tilde k_x.k_y)|^2,
\end{alignat}
where the normalization factor is due to the multiplicity associated with the total flux in Eq. (\ref{TotPumCha2}).
It follows from Eq. (\ref{TotPumCha2}) that 
the  normalized center of mass gives the second Chern number
\begin{alignat}1
c_2=-\mbox{sgn}(e)
\left[\mbox{ sum of discontinuities of }p_z(\theta)\right].
\end{alignat}

In Fig. \ref{f:com}, we show the spectral flow as a function of $\theta$ and the normalized center of mass $p_z(\theta)$.
In both cases in Fig. \ref{f:com}, the structure of the vacuum (negative energy states)
changes at $\theta=0$, and shows the discontinuity in 
the center of mass, from which the second Chern number can be obtained. 
The result is consistent with Eq. (\ref{ExpCheNum2}).

\begin{figure}[htb]
\begin{center}
\begin{tabular}{c}
\includegraphics[scale=0.8]{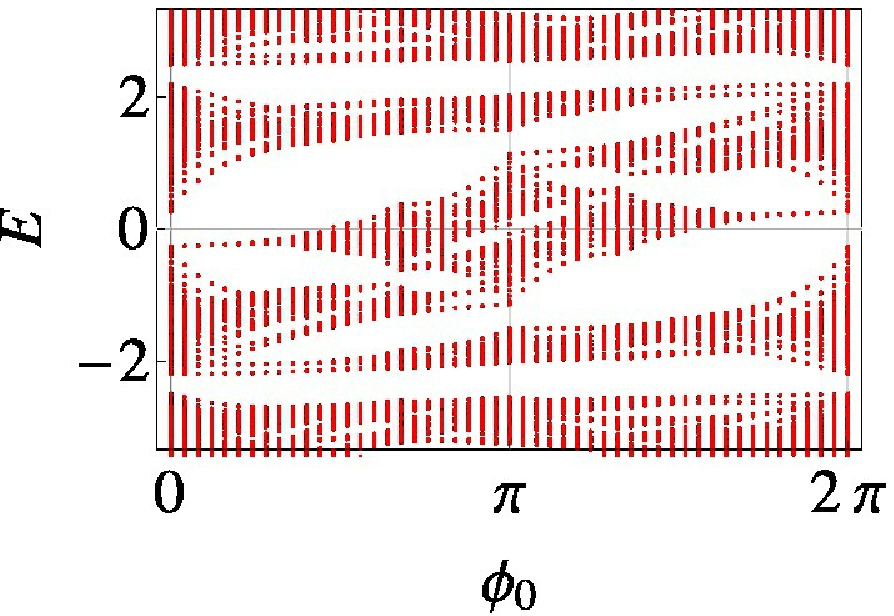}\\
\hspace*{3mm}
\includegraphics[scale=0.8]{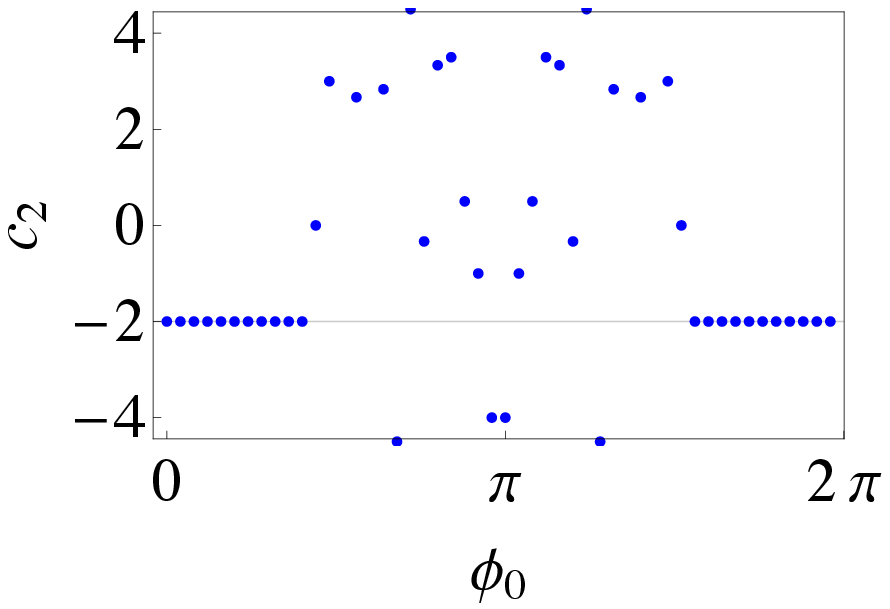}\\
\end{tabular}
\caption{
Upper figure shows the Hofstadter diagram of the Hamiltonian (\ref{ModWDHam}) 
as a function of the magnetic flux per plaquette (\ref{FluPla}). 
The parameters are $q=50$, $N_y=N_z=6$, $\tilde q=6$, and $\tilde p=1$ is fixed, 
whereas $p$ is changed by $\Delta p=1$.
The model parameter is $m=3b$.
Lower figure  shows the difference of the number of the negative energy states defined in the rhs of Eq. (\ref{StrFor}).
}
\label{f:asy}
\end{center}
\end{figure}

\subsection{Generalized Streda formula}
\label{s:Streda}

For reference, we here present a method of computing the second Chern number based on 
the generalized Streda formula.\cite{Fukui:2016aa}
Without an electro-magnetic field, the Wilson-Dirac Hamiltonian (\ref{WDHam}) in the momentum space reads
\begin{alignat}1
{\cal H}(k)=&\Gamma^j\sin k_j+m\Gamma^4\sin k_4
\nonumber\\
&-\Gamma_5\left[m\cos k_4+b\sum_{j=1}^3(\cos k_j-1)\right],
\end{alignat}
where we have introduced $k_4=\theta$, and 
new hermitian $\Gamma$ matrices $\Gamma^j=\alpha^j$ ($j=1,2,3$), $\Gamma^4=-i\beta\gamma_5$, and $\Gamma^5=\beta$,
with $\{\Gamma^\mu,\Gamma^\nu\}=2\delta^{\mu\nu}$.
Now let us regard $k_4$ as the frequency of discrete imaginary time.
Then, we can reconstruct an equivalent lattice fermion such that
\begin{alignat}1
{\cal H}=&\Gamma^j D_j^{\rm L}+m\Gamma^4 D_4^{\rm L}
-\Gamma^5\left[\frac{m}{2}(\Delta_4^{\rm L}+2)+\frac{ba}{2}\Delta^{\rm L}\right],
\label{ModWDHam}
\end{alignat}
where $D_4^{\rm L}\equiv (\nabla_4+\nabla_4^*)/2$ and $\Delta_4^{\rm L}=(\nabla_4-\nabla_4^*)/a$
are defined for the new coordinate $x^4$ in the same way as Eqs. (\ref{LatDirOpe}) 
and (\ref{CovDifOpe}).

Using this four dimensional Hamiltonian, one can define the overlap Dirac operator,
\cite{Neuberger:1998ab,Neuberger:1998aa} which obeys the Ginsparg-Wilson relation. \cite{Ginsparg:1982aa}
This enables us to define the chiral anomaly on the lattice. \cite{Luscher:1998aa,Kikukawa:1999aa,Suzuki01071999}
Taking the continuum limit, it indeed 
reproduces the chiral anomaly in arbitrary dimensions
with  a nontrivial Chern number as a coefficient.\cite{1126-6708-2002-09-025}
It has been shown that the chiral anomaly thus obtained 
is given by the spectral asymmetry of the above Hamiltonian, and hence one can compute the second Chern number 
from the spectral flow of the Hamiltonian. \cite{Fukui:2016aa}
This is referred to as the generalized Streda formula.
In what follows,  we use $N_-$ rather than the spectral asymmetry $\eta=(N_+-N_-)/2$, where
$N_\pm$ is the number of positive and negative energy states.

Consider the system which includes a static and uniform magnetic field in the $x^3=z$ direction, as 
studied in Secs. \ref{s:3dpump}, \ref{s:1dpump}, and \ref{s:BEC}, and 
introduce a fictitious electric field associated with the imaginary time $x^4$ direction.
As shown in Ref. \cite{Fukui:2016aa}, the density of occupied (negative energy) states 
\begin{alignat}1
n_-=\frac{1}{2V}{\rm Tr }\left(1-\frac{{\cal H}}{\sqrt{\cal H}^2}\right)=\frac{N_-}{V}
\end{alignat}
of the Hamiltonian (\ref{ModWDHam}) as a function of the magnetic field yields the second Chern number,
\begin{alignat}1
\frac{\partial n_-}{\partial (\bm B\cdot\bm E)}=-\frac{c_2}{(2\pi)^2},
\end{alignat}
where $\bm E$ is the fictitious electric field.
To be concrete, assume that the magnetic field is included in the Landau gauge in Sec. \ref{s:1dpump}.
The fictitious electric field is also included in the $z$ direction such that
\begin{alignat}1
e\bm A(\bm x)=(0,\mbox{sgn(}e) \phi\frac{x}{a},\mbox{sgn(}e) \tilde\phi\frac{x^4}{a}),
\label{LanGau_2}
\end{alignat}
where $\tilde \phi=2\pi \tilde p/\tilde q$, gives the fictitious electric field $(0,0,E)$ such that
$E=-\partial A_z/\partial x^4=-\mbox{sgn}(e)\tilde\phi/a^2$.
For numerical calculations, we set the system size as $q$, $N_y$, $N_z$, and $\tilde q$ sites for
$x$, $y$, $z$, and $x^4$ directions, respectively, with periodic boundary conditions imposed in all directions.
Then, the volume of the system is $V=qN_yN_z\tilde qa^4$, and therefore,
\begin{alignat}1
c_2=-(2\pi)^2\frac{\Delta N_-}{qN_yN_z\tilde qa^4\Delta(BE)}=\frac{\Delta N_-}{N_yN_z\tilde p\Delta p},
\label{StrFor}
\end{alignat}
where $\Delta p=1$, and we have assumed that the electric field is fixed.
In Fig. \ref{f:asy}, we show the spectrum as a function $\phi_0$, and corresponding difference of 
the density of the occupied states. 
This figure tells that the the second Chern number is $c_2=-2$, 
consistent with the previous results.
In passing, we mention that in the case $m=b$, we can also reproduce $c_2=1$ in the same manner.

The Hofstadter diagram in Fig. \ref{f:asy} is different from that in Fig. \ref{f:but}, 
since the former includes a finite fictitious electric field. 
In other words, such a difference enables us to compute the second Chern number. 
Without the fictitious electric field, the anomaly is trivially vanishing.

\section{Summary and discussion}
\label{s:sd}

In summary, we studied mainly the 3D topological pump analytically and numerically in detail in this paper. 
We introduced a variant of the Wilson-Dirac model defined
on the spatial lattice but in continuous time, including two kinds of mass terms depending generically on 
$\bm x$ as well as $t$.  
We derived the conserved current density on the lattice and calculate it in the continuum limit,
or in other words in the small field limit.
For the temporal pump, the result was checked by numerical calculations from various methods
as follows.

Firstly, the 3D pump governed by the second Chern number (\ref{Qz1}) or (\ref{TotPumCha2})
can be viewed as a set of 1D pump described by the first Chern number (\ref{1ChaNum}) or (\ref{Qz2}).
It should be noted that the latter description is valid even in a strong magnetic field as long as the 
mass gap is open.
Both results lead to the relationship between two Chern numbers (\ref{RelC1C2}).
We showed  by the numerical calculation of the first Chern number using the Berry curvature of the wave functions
that Eq. (\ref{RelC1C2}) is valid in a strong magnetic field regime up to the gap closing point. 
It would be an interesting problem to ask whether the relationship (\ref{RelC1C2}) is restricted only to the 
present system or more applicable to other cases. 
Since the second Chern number is generically due to
non-Abelian Berry curvature, its numerical calculation is very hard, and therefore, 
a simple relationship like (\ref{RelC1C2}) is quite helpful.

Secondly, as the Chern number description of 
the number of the pumped particles is for the bulk system, 
the bulk-edge correspondence enables us to 
observe the 3D pump as the flow of the surface states.
We showed that the bulk-edge correspondence established in a 1D pump \cite{Hatsugai:2016aa}
can be applied to the 
present 3D system, and discontinuities of the center of mass of the occupied particles,
which are the contribution from the surface states, reproduces the 
correct number of the pumped particles.
The center of mass is one of important observables for the topological pump: \cite{Wang:2013fk_pump}
Indeed,  in the recent experiments of the 1D pump, \cite{Nakajima:2016aa,Lohse:2016aa}
the center of mass played a central role.
Thus, it would be expected that 3D pump can be detected experimentally using the center of mass of the occupied particles.
This would also imply the experimental observation of the chiral anomaly. 

In passing, we would like to add a comment on the effect of interactions.
So far we have discussed the bulk-edge correspondence for a particle pump of a noninteracting system by 
observing the discontinuities of the center of mass of the ground state {\it in contact with a particle reservoir}.  
Since the present pumping is of topological origin, small interactions cannot change the quantized discontinuities of the 
center of mass.
On the other hand, for systems with strong interactions, fractional pumping have been proposed
for 1D systems with degenerate ground states.
\cite{Zeng:2016aa,Li:2017aa,Taddia:2017aa}
In such cases, the discontinuities of the center of mass is not obvious, but if we introduce a particle reservoir also for such systems
and consider the ground states with different number of particles, we could discuss the discontinuities for the degenerate ground states.
This is because of the universality of the bulk-edge correspondence: 
The bulk topological properties should be closely related with the edge states also for interacting systems.

Thirdly, we applied the generalized Streda formula, which is based on the chiral anomaly of the Dirac fermion,
to compute the second Chern number.
Let us here mention the anomaly of the present system.
The current (\ref{ChaDen3}) may be derived from the effective action, as has been done in Ref. \cite{Qi:2008aa}. 
Although we did not directly calculate the effective action, 
we can derive it from the expressions of the current (\ref{ChaDen1}) and (\ref{ChaDen3}) as follows:
Let $\Gamma_{\rm eff}[\theta,A]$ be the effective action defined by 
\begin{alignat}1
i\Gamma_{\rm eff}[\theta,A]=\ln{\rm Det}\left[(i\slashed D-me^{i\gamma_5\theta})/(i\slashed D-m)\right].
\end{alignat}
Then, the current can be obtained by
\begin{alignat}1
\langle j^\mu(x)\rangle=\frac{1}{e}\frac{\delta \Gamma_{\rm eff}[\theta,A]}{\delta A_\mu(x)},
\end{alignat}
from which  we have for $d=1+1$ system
\begin{alignat}1
\Gamma_{\rm eff}[\theta,A]=\frac{e}{2}\int d^2x\epsilon^{\mu\nu}P_1(\theta)F_{\mu\nu},
\end{alignat}
and for $d=3+1$ system
\begin{alignat}1
\Gamma_{\rm eff}[\theta,A]=-\frac{e^2}{16\pi}\int d^4x\epsilon^{\mu\nu\rho\sigma}P_3(\theta)
F_{\mu\nu}F_{\rho\sigma},
\end{alignat}
where the charge polarization ($d=1+1$) or the magneto-electric polarization ($d=3+1$), $P_{d-1}(\theta)$,
is defined by \cite{Qi:2008aa}
\begin{alignat}1
P_{d-1}(\theta)=\int_0^\theta d\theta G_{d-1}(\theta).
\end{alignat}
For $\theta=2\pi$, the effective action gives the chiral anomaly. 
On the other hand, in Sec. \ref{s:Streda} in this paper, 
we demonstrated the manifestation of the anomaly
based on a related method developed in Ref. \cite{Fukui:2016aa}. 
Namely, we calculated the second Chern number  from spectral asymmetry of the four dimensional Hamiltonian with an electric field as well as a magnetic field. As shown in Ref. \cite{Fukui:2016aa}, the spectral asymmetry 
gives the chiral anomaly of the  overlap Dirac operator \cite{Neuberger:1998ab,Neuberger:1998aa} 
obeying the Ginsparg-Wilson relation.
It may be usually natural to use the Wilson-Dirac Hamiltonian to construct the overlap operator
$D=\frac{1}{a}\left(1-\gamma_5\frac{\cal H}{\sqrt{{\cal H}^2}}\right)$.
However, for any gapped Hamiltonian, the overlap operator obeys the Ginsparg-Wilson relation.\cite{Ginsparg:1982aa}
Thus, it would be an interesting future problem to seek the possible Hamiltonian for the overlap operator.

Finally, we would like to add a comment on recent observations concerning
the four dimensional (4D) topological pump. 
In Refs. \cite{1705.08361,1705.08371}, the well-known two dimensional (2D) (or $d=1+1$) topological pump models 
such as the Harper pump model \cite{Hatsugai:2016aa}
or Rice-Mele model \cite{Rice:1982qf,Xiao:2010fk}
$H(k,t)$ are extended to a 4D model considering the direct sum, $H(k_1,t)+H'(k_2,s)$. 
This allows a simple relationship between the  
second Chern number that governs the topological properties of a 4D system
and the first Chern number of each 2D (or $d=1+1$) subsystem.
In spite of some weak couplings between two subsystems in the experimental setup, the expected second Chern number 
has been observed indeed. 
However, if the couplings become larger, the topological change may be expected, which is an interesting future issue to explore. 
To this end, we note that 
for the lowest non-degenerate band studied in Refs. \cite{1705.08361,1705.08371}, the second Chern number 
associated with the U(1) Berry curvature can be computed directly on the lattice. 
\cite{Luescher:1999uq,Luescher:1999fk,1126-6708-2002-09-025}
Also it may be interesting to develop several numerical techniques studied in Sec. \ref{s:MEP} for these non-Dirac systems, or
to apply the techniques of the entanglement Chern number, which can separate the Chern number into 
those of subsystems.\cite{Fukui:2014qv,Araki:2017aa}.

\section*{Acknowledgments}

We would like to thank Y. Hatsugai for fruitful discussions.
This work was supported in part by Grants-in-Aid for Scientific Research Numbers 17K05563 and 17H06138 from
the Japan Society for the Promotion of Science.

\appendix

\section{Current density in the continuum limit}
\label{s:CurDen}

In this Appendix, we calculate the current density
given by Eq. (\ref{WilDirCur}).
It is similar to the charge density, but it includes the lattice corrections.
From Eq. (\ref{WilDirCur}), we need to calculate
\begin{widetext}
\begin{alignat}1
\langle j^l(x)\rangle&=
-\lim_{x'\rightarrow x}
\tr \langle0|
T\Big\{\gamma^l\psi(x)\bar\psi(x')
+\frac{a}{2}\gamma^l\left[\nabla_l\psi(x)\bar\psi(x')
+\psi(x)\bar\psi(x')\overleftarrow{\nabla}_l'\right]
+\frac{iba}{2}\left[\nabla_l\psi(x)\bar\psi(x')
-\psi(x)\bar\psi(x')\overleftarrow{\nabla}_l'\right]\Big\}|0\rangle,
\end{alignat}
where the repeated $l$ in the middle term is not summed, as have been noticed.
Using the propagator (\ref{WilDirPro}), this can be written as 
\begin{alignat}1
\langle j^l(x)\rangle
&=\frac{1}{a^{d-1}}
\int_{-\infty}^\infty\frac{d\omega}{2\pi i}
\int\frac{d^{d-1}k}{(2\pi)^{d-1}}
e^{-i\frac{kx}{a}}
\tr\Bigg\{\left(\gamma^l+\frac{a}{2}\gamma^l\nabla_l+\frac{iba}{2}\nabla_l\right)
\frac{1}{ia\slashed D^{\rm L}-me^{i\gamma_5\theta}-\frac{b}{2}a^2\Delta^{\rm L}+i\epsilon}
\nonumber\\
&\qquad+\frac{1}{ia\slashed D^{\rm L}-me^{i\gamma_5\theta}-\frac{b}{2}a^2\Delta^{\rm L}+i\epsilon}
\left(\frac{a}{2}\gamma^l\overleftarrow{\nabla}_l-\frac{iba}{2}\overleftarrow{\nabla}_l\right)
\Bigg\}e^{i\frac{kx}{a}} .
\label{Jl_1}
\end{alignat}
Note that in the limit $a\rightarrow 0$, the difference $\overleftarrow{\nabla}_l$ becomes 
\begin{alignat}1
e^{-i\frac{kx}{a}} a\overleftarrow{\nabla}_j e^{i\frac{kx}{a}}=e^{ik_j}a\overleftarrow{\nabla}_j+e^{ik_j}-1
=-e^{ik_j}aD_j+e^{ik_j}-1+O(a^2).
\end{alignat}
Using this, together with (\ref{DiraOpeCon_1}), we have
\begin{alignat}1
\langle j^l(x)\rangle
=&\frac{1}{a^{d-1}}
\int_{-\infty}^\infty\frac{d\omega}{2\pi i}
\int\frac{d^{d-1}k}{(2\pi)^{d-1}}
\tr(\gamma^lc_l+bs_l)
\left(\slashed{s}-me^{-i\gamma_5\theta}-bc_{\rm s}\right)
\nonumber\\
&\times\frac{1}
{
\mu^2-s^2+me^{i\gamma_5\theta}\gamma_5a \tilde{\slashed{\partial}}\theta
-mbe^{-i\gamma_5\theta}\gamma_5a\tilde\partial_{\rm s}\theta
-\frac{i\gamma^\rho\gamma^\sigma}{2}ea^2\widetilde F_{\rho\sigma}
+ib\gamma^\rho ea^2\widetilde F_{\rho,{\rm s}}
},
\label{Jl_2}
\end{alignat}
where the repeated $l$ in the rhs is not summed, 
and we have already omitted irrelevant operators after the limit 
$a\rightarrow0$ as well as after the trace over the $\gamma$ matrices.
Expanding the propagator in (\ref{Jl_2}) with respect to $a$ as we did to compute the charge density in Sec. \ref{s:ChaDen},
we can calculate the current density.
In what follows, we briefly show several steps of the calculations separately in $d=1+1$ and $d=3+1$.

\subsection{$d=1+1$ system}

Corresponding to Eq. (\ref{ChaDen1Ori}), the following expression can be obtained for the current density,
\begin{alignat}1
\langle j^1(x)\rangle
=&\frac{1}{a}
\int_{-\infty}^\infty\frac{d\omega}{2\pi i}
\int\frac{dk_1}{2\pi}
\frac{
\tr(\gamma^1c_1+bs_1)
\left(\slashed{s}-me^{-i\gamma_5\theta}-bc_{\rm s}\right)(-ma\gamma_5)
(e^{i\gamma_5\theta}\tilde{\slashed{\partial}}\theta
-be^{-i\gamma_5\theta}\tilde\partial_{\rm s}\theta)
}
{(\mu^2-s^2-i\epsilon)^2}.
\label{Jl_d1}
\end{alignat}
In the numerator, the terms which survive after the trace and integration over $\omega$ are
\begin{alignat}1
(-ma)\left[
\tr\gamma_5\gamma^1c_1(me^{-i\gamma_5\theta}+bc_{\rm s})e^{i\gamma_5\theta}\tilde{\slashed\partial}\theta
- bs_1\tr\gamma_5\slashed se^{i\gamma_5\theta}\tilde{\slashed\partial}\theta
\right]
=2am\left[(m+bc_{\rm s}\cos\theta)c_1+bs_1^2\cos\theta\right]\epsilon^{10}\partial_0\theta .
\end{alignat}
This is nothing but Eq. (\ref{ChaDen1}) for $\mu=1$.
Thus, we have established that the result in Eq. (\ref{ChaDen1}) is valid for any $\mu$.

\subsection{$d=3+1$ system}

From Eq. (\ref{Jl_2}), we obtain the following equation similar to Eq. (\ref{J0_3}),
\begin{alignat}1
\langle j^l(x)\rangle
&=\frac{1}{a^3}
\int_{-\infty}^\infty\frac{d\omega}{2\pi i}
\int\frac{d^3k}{(2\pi)^3}
\frac{1}{(\mu^2-s^2-i\epsilon)^3}
\nonumber\\
&\times
\tr(\gamma^lc_l+bs_l)\left(\slashed{s}-me^{-i\gamma_5\theta}-bc_{\rm s}\right)
\left(
-me^{i\gamma_5\theta}\gamma_5a \tilde{\slashed{\partial}}\theta
+mbe^{-i\gamma_5\theta}\gamma_5a\tilde\partial_{\rm s}\theta
+\frac{i}{2}\gamma^\rho\gamma^\sigma ea^2\widetilde F_{\rho\sigma}
-ib\gamma^\rho ea^2\widetilde F_{\rho,{\rm s}}
\right)^2 .
\end{alignat}
As in the case of the charge density, the product terms between $\partial\theta$ and $\widetilde F$ survive 
in the limit $a\rightarrow0$ and after the trace over the $\gamma$ matrices. To be concrete, the trace for the $\gamma$
matrices yields
\begin{alignat}1
ea^3\Big\{&
\tr\gamma^lc_l(-me^{-i\gamma_5\theta}-bc_{\rm s})(-me^{i\gamma_5\theta}\gamma_5a\tilde{\slashed\partial}\theta)
(i\gamma^\rho\gamma^\sigma \widetilde F_{\rho\sigma})
+\tr\gamma^lc_l\slashed smbe^{-i\gamma_5\theta}\gamma_5\tilde\partial_{\rm s}\theta
(i\gamma^\rho\gamma^\sigma \widetilde F_{\rho\sigma})
\nonumber\\
&+2\tr\gamma^lc_l\slashed s(-me^{i\gamma_5\theta}\gamma_5\tilde{\slashed\partial}\theta)
(-ib\gamma^\rho\widetilde F_{\rho,{\rm s}})
+bs_l\tr\slashed s(-me^{i\gamma_5\theta}\gamma_5\tilde{\slashed\partial}\theta)
(i\gamma^\rho\gamma^\sigma\widetilde F_{\rho\sigma})
\Big\}
\nonumber\\
=&-mea^3i\Big\{
(m+bc_{\rm s}\cos\theta)c_l\tr\gamma_5\gamma^l
(\tilde{\slashed\partial}\theta)
\gamma^\rho\gamma^\sigma \widetilde F_{\rho\sigma}
-b\cos\theta c_l\tr\gamma_5\gamma^l\slashed s\tilde\partial_{\rm s}\theta
\gamma^\rho\gamma^\sigma \widetilde F_{\rho\sigma}
\nonumber\\
&
-2b\cos\theta c_l\tr\gamma_5\gamma^l\slashed s\tilde{\slashed\partial}\theta
\gamma^\rho\widetilde F_{\rho,{\rm s}}
-bs_l\cos\theta\tr\gamma_5\slashed s\tilde{\slashed\partial}\theta\gamma^\rho\gamma^\sigma\widetilde F_{\rho\sigma}
\Big\}
\nonumber\\
=&-4ea^3m\left[m+b\sum_{j=1}^3(\sec k_j-1)\cos\theta\right]\left(\prod_{j=1}^3\cos k_j\right)
\epsilon^{l\nu\rho\sigma}(\partial_\nu\theta) F_{\rho\sigma} .
\end{alignat}
\end{widetext}
Thus, we finally reach Eq. (\ref{ChaDen3}) for $\mu=l$, and the formula (\ref{ChaDen3}) has been established 
for any $\mu$.

\section{Mathematical formulas}
In this Appendix, we show some mathematical formulas used in the text.
\subsection{Trace for $\gamma$ matrices}
The trace of the $\gamma$ matrices including $\gamma_5$ is summarized as follows:
In $d=1+1$ dimensions, 
\begin{alignat}1
\tr\gamma_5=\tr \gamma_5\gamma^\mu=0,\quad
\tr \gamma_5\gamma^\mu\gamma^\nu=-2\epsilon^{\mu\nu} .
\label{TraGam1D}
\end{alignat}
In $d=3+1$ dimensions, 
\begin{alignat}1
&\tr\gamma_5=\tr \gamma_5\gamma^\mu=\tr \gamma_5\gamma^\mu\gamma^\nu=\tr \gamma_5\gamma^\mu\gamma^\nu\gamma^\rho=0,  
\nonumber\\ 
&\tr \gamma_5\gamma^\mu\gamma^\nu\gamma^\rho\gamma^\sigma=-4i\epsilon^{\mu\nu\rho\sigma}.
\label{TraGam3D}
\end{alignat}

\subsection{Integral}

For the $d$ dimensional momentum integration in the continuum model, we use
\begin{alignat}1
\int\frac{d^dk}{i(2\pi)^d}\frac{1}{(1-k^2-i\epsilon)^n}=\frac{\Gamma(n-d/2)}{(4\pi)^{d/2}\Gamma(n)}.
\label{Int_1}
\end{alignat}
As to the integration over $\omega$ in the lattice model,
\begin{alignat}1
I_n\equiv\int_{-\infty}^\infty\frac{d\omega}{2\pi i}\frac{1}{(\Omega^2-\omega^2-i\epsilon)^n}.
\end{alignat}
we simply have in the $n=1$ case,
\begin{alignat}1
I_1=\frac{1}{2\Omega}.
\end{alignat}
Then,  we have
\begin{alignat}1
&I_2=-\frac{d}{d\Omega^2}I_1=\frac{1}{4\Omega^3}, 
\label{IntOme1}
\\
&I_3=-\frac{1}{2}\frac{d}{d\Omega^2}I_2=\frac{3}{16\Omega^3}.
\label{IntOme2}
\end{alignat}


\end{document}